\documentclass[11pt, draftcls, journal, onecolumn]{IEEEtran}
\usepackage{amsmath}
\usepackage{amsfonts}
\usepackage{amssymb}
\usepackage{setspace}
\usepackage{float}
\usepackage{graphicx}
\usepackage{amsthm}
\usepackage{color}
\usepackage{cite}
\usepackage{bm}
\usepackage{epsfig,psfrag}
\usepackage{multirow}
\usepackage[yyyymmdd,hhmmss]{datetime} % gives time of the compiling
\usepackage{epsfig}
\usepackage{psfrag}
\usepackage{pst-node}
\usepackage{fancyhdr,bbm}
\usepackage{color}
\usepackage{wrapfig}
\usepackage{comment}
\usepackage{booktabs}

\usepackage[columnwise,switch,mathlines]{lineno} % this adds line numbers for proof reading 
\newcommand{\be}{\begin{equation}}
\newcommand{\ee}{\end{equation}}

\newcommand{\T}{\text{T}}

\allowdisplaybreaks

\newlength{\figureheight}
\newlength{\figurewidth}
\definecolor{temporalgreen}{RGB}{0,128,0}

\newcommand{\V}[1]{\bm{#1}}

\definecolor{BLUE}{rgb}{0,0,1}

\newdateformat{monthyeardate}{\monthname[\THEMONTH] \THEDAY, \THEYEAR} % specifies the format of the date

\newcommand{\paperTitle}{Space-based Global Maritime Surveillance. Part II: Artificial Intelligence and Data Fusion Techniques\vspace{1.2mm}}

\floatstyle{boxed}
\newfloat{floatbox}{thp}{lob}[section]
\floatname{floatbox}{Text Box}

\excludecomment{comment}

\begin{document}

\renewcommand{\baselinestretch}{1.4}\small\normalsize

\title{\paperTitle}

\author{Giovanni~Soldi,
		Domenico~Gaglione,~\IEEEmembership{Member,~IEEE},
		Nicola~Forti,
		Alessio~Di~Simone,~\IEEEmembership{Member,~IEEE},
		Filippo~Cristian~Daffin\`{a},
		Gianfausto~Bottini,
		Dino~Quattrociocchi,
		Leonardo~M.~Millefiori,~\IEEEmembership{Member,~IEEE},
		Paolo~Braca,~\IEEEmembership{Senior Member,~IEEE},
		Sandro~Carniel,
		Peter~Willett,~\IEEEmembership{Fellow,~IEEE},
		Antonio~Iodice,~\IEEEmembership{Senior Member,~IEEE},
		Daniele~Riccio,~\IEEEmembership{Fellow,~IEEE}, and 
		Alfonso~Farina~\IEEEmembership{Life Fellow,~IEEE}\vspace{-8mm}\thanks{G.\ Soldi, D.\ Gaglione, N.\ Forti, L.\ M.\ Millefiori, P.\ Braca and S.\ Carniel are with the NATO Centre for Maritime Research and Experimentation (CMRE), La~Spezia, Italy (e-mail: giovanni.soldi@cmre.nato.int, domenico.gaglione@cmre.nato.int, nicola.forti@cmre.nato.int, leonardo.millefiori@cmre.nato.int, paolo.braca@cmre.nato.int, sandro.carniel@cmre.nato.int). A.\ Di Simone, A.\ Iodice and D.\ Riccio are with the Department of Electrical Engineering and Information Technology (DIETI) of the university of Naples Federico II, Italy (e-mail: alessio.disimone@unina.it, antonio.iodice@unina.it, daniele.riccio@unina.it). F.\ C.\ Daffin\`{a}, G.\ Bottini and D.\ Quattrociocchi are with E-GEOS in Rome, Italy (e-mail: filippo.daffina@e-geos.it, gianfausto.bottini@e-geos.it, dino.quattrociocchi@e-geos.it). 
P.\ Willett is with the Department of Electrical and Computer Engineering, University of Connecticut, Storrs, CT 06269-2157 USA (e-mail: peter.willett@uconn.edu). P.\ Willett was supported in part by NIUVT and by AFOSR under contract FA9500-18-1-0463. A.\ Farina is a professional consultant in Rome, Italy (e-mail: alfonso.farina@outlook.it).
This work was supported in part by the NATO Allied Command Transformation under project SAC000A08.
%by the European Research Council (ERC) under grant 700478 (project RANGER) within the Horizon 2020 program,
%% the European UnionÕs Horizon 2020 research and innovation programme
%% within project RANGER (grant agreement no. 700478),
%% \bl{Project Data Knowledge Operational Effectiveness (DKOE) - SAC000808,} 
%by the Austrian Science Fund (FWF) under grants J3886-N31 and P27370-N30, and by the Czech Science Foundation (GA\v{C}R) under grant 17-19638S.
} 
		}
		
\maketitle

\begin{abstract}
\label{sec:abstract}
Maritime surveillance (MS) is of paramount importance for search and rescue operations, fishery monitoring, pollution control, law enforcement, migration monitoring, and national security policies. 
Since ground-based radars and automatic identification system (AIS) do not always provide a comprehensive and seamless coverage of the entire maritime domain, the use of space-based sensors is crucial to complement them. We reviewed space-based
technologies for MS in the first part of this work, titled ``Space-based Global Maritime Surveillance. Part I: Satellite Technologies''~\cite{SpaceAESM_1:J20}. 
However, future MS systems combining multiple terrestrial and space-based sensors with additional information sources will require dedicated artificial intelligence and data fusion techniques for the processing of raw satellite images and fuse heterogeneous information. The second part of our work focuses on the most promising artificial intelligence and data fusion techniques for MS using space-based sensors.
\end{abstract}

\section{Introduction}
\label{sec:introduction}
Maritime surveillance (MS) aims at providing seamless wide-area operational pictures in coastal areas and the oceans in real time. Multiple heterogeneous sensors and information sources are nowadays available for MS, all with their advantages and limitations, and significant research and engineering effort has been made for their combination. 
Besides the most common terrestrial sensors for MS, e.g., the automatic identification system~(AIS), X-band radars, and over-the-horizon (OTH) radars, space-based sensor technologies  enable persistent monitoring of the maritime domain and ship traffic on a global scale even in remote areas of the Earth. 
Space-based remote sensing technologies, reviewed in the first part of this work~\cite{SpaceAESM_1:J20}, include satellite-based AIS (Sat-AIS), synthetic aperture radar (SAR), multi-spectral (MSP) and hyper-spectral (HSP) optical sensors and global navigation satellite systems reflectometry (GNSS-R).
Space-based sensors for Earth observation installed on satellites allow collecting images of very large areas in remote regions of the globe with relatively short latency, and hence are strongly relevant to MS. 

As a consequence of the deployment and spread of space-based sensor technologies, advanced data
processing paradigms, e.g., big data analysis, machine learning, artificial intelligence (AI) and data fusion, are now extremely needed to fully exploit the 
wide availability of large data sets of satellite images. 
In particular, the development of future MS systems combining multiple sensors requires dedicated algorithms to process raw satellite images, detect and classify ships and fuse information from heterogeneous sensors. These algorithms support MS by processing and organizing the increasing amount of heterogeneous information. 
The extracted data and readily-understandable information digested therefrom will help end-users, such as governmental and military authorities, coast guards, and
police, to detect anomalies, threats such as oil spills, piracy, and human trafficking, and act in time to prevent accidents and wrongdoing.
In this second part of our work, we describe the main AI and imaging techniques for image segmentation, target detection and classification, and provide possible use cases with real images acquired by satellite sensors. 
Then, we describe recent Bayesian and statistical fusion techniques to extract knowledge from the troves of historical Sat-AIS data, such as most common maritime routes, and to track multiple targets by fusing information collected by multiple heterogeneous sensors. Among these, multitarget tracking (MTT) algorithms based on the sum-product algorithm (SPA) are gaining popularity thanks to their ability to fuse information from heterogeneous sources, to their scalability, i.e., low computational complexity in terms of the
number of information sources, targets and measurements, and to their capability to include contextual information, e.g., maritime routes and ships class information extracted from satellite images. We also provide a use case that confirms the strength of SPA-based MTT algorithms when combined with information acquired by satellite sensors is also provided.

\section{Advanced AI Techniques for Satellite Sensor Data}
\label{sec:Advanced_AI}
\begin{figure}[!t]
\centering
\includegraphics[width=0.5\columnwidth]{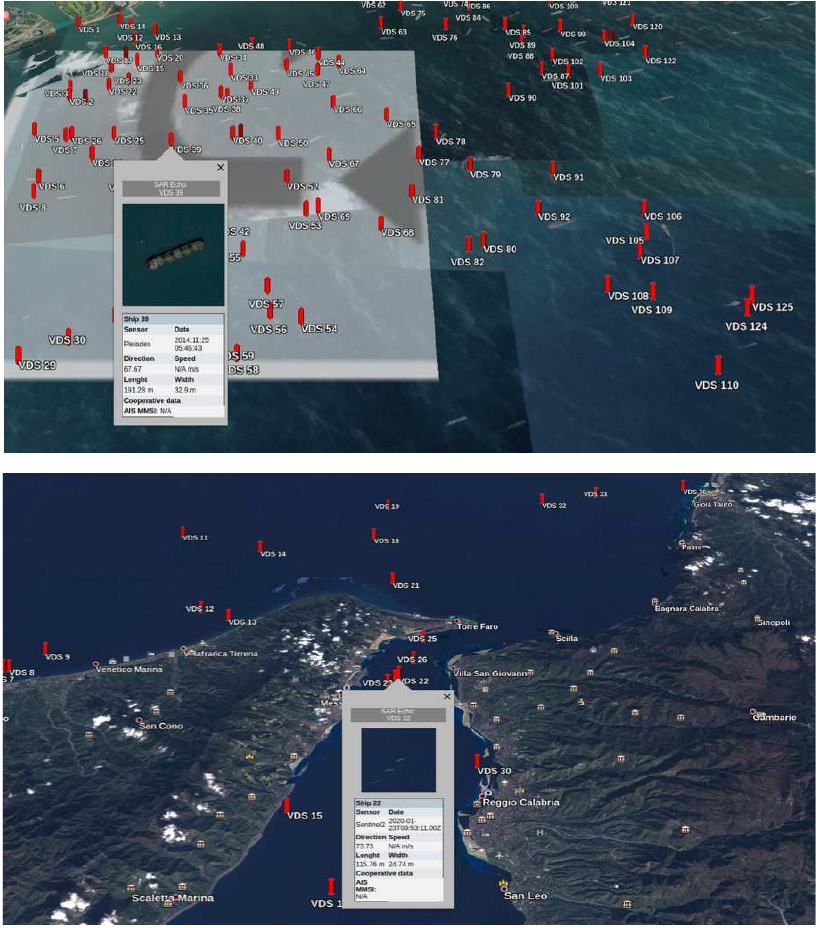}
\caption{Examples of satellite images augmented by a digital report for each detected ship. 
(Top) MSP sensor: Pleiades, Area: close to Khark island (Iran), Date: 25th November 2014. (Bottom) MSP sensor: Sentinel-2, City: Strait of Messina (Italy), Date: 23rd January 2020.
}
\label{fig:digital_report}
\vspace{-1mm}
\end{figure}
The proliferation of space-based sensors collecting images of the Earth has called for the development of advanced AI and imaging techniques to extract ship features and identities to improve the surveillance capability.   
In particular, and analogous to many other scientific fields, deep neural network (DNN) techniques for detection and segmentation of specific targets from satellite images have recently received increasing interest. %in the maritime surveillance community . 
%Since few years, neural network (CNN) architectures have increased their importance in object segmentation from satellite images being now crucial for defense, security, maritime domain awareness  and similar fields largely improving Computer Vision techniques accuracies.
%The main purpose 
Advanced AI and DNN techniques allow, for example, the production of a ``digital report'' for any satellite image  by performing vessel detection, segmentation, 
classification, and identification: by estimating ships dimensions and headings, and by providing geographic coordinates and timestamp. 
Fig.~\ref{fig:digital_report} shows some examples of satellite images augmented by a digital report for each detected ship.

In this section, we provide a review of the main AI and statistical techniques for target detection, segmentation and classification in images acquired by satellite sensors.

\vspace{-2mm}
\subsection{DNN Architectures for Satellite Image Segmentation and Target Classification }
\label{subsec:main_architectures_obj_class}

\begin{figure}[!t]
\centering
\includegraphics[width=0.9\columnwidth]{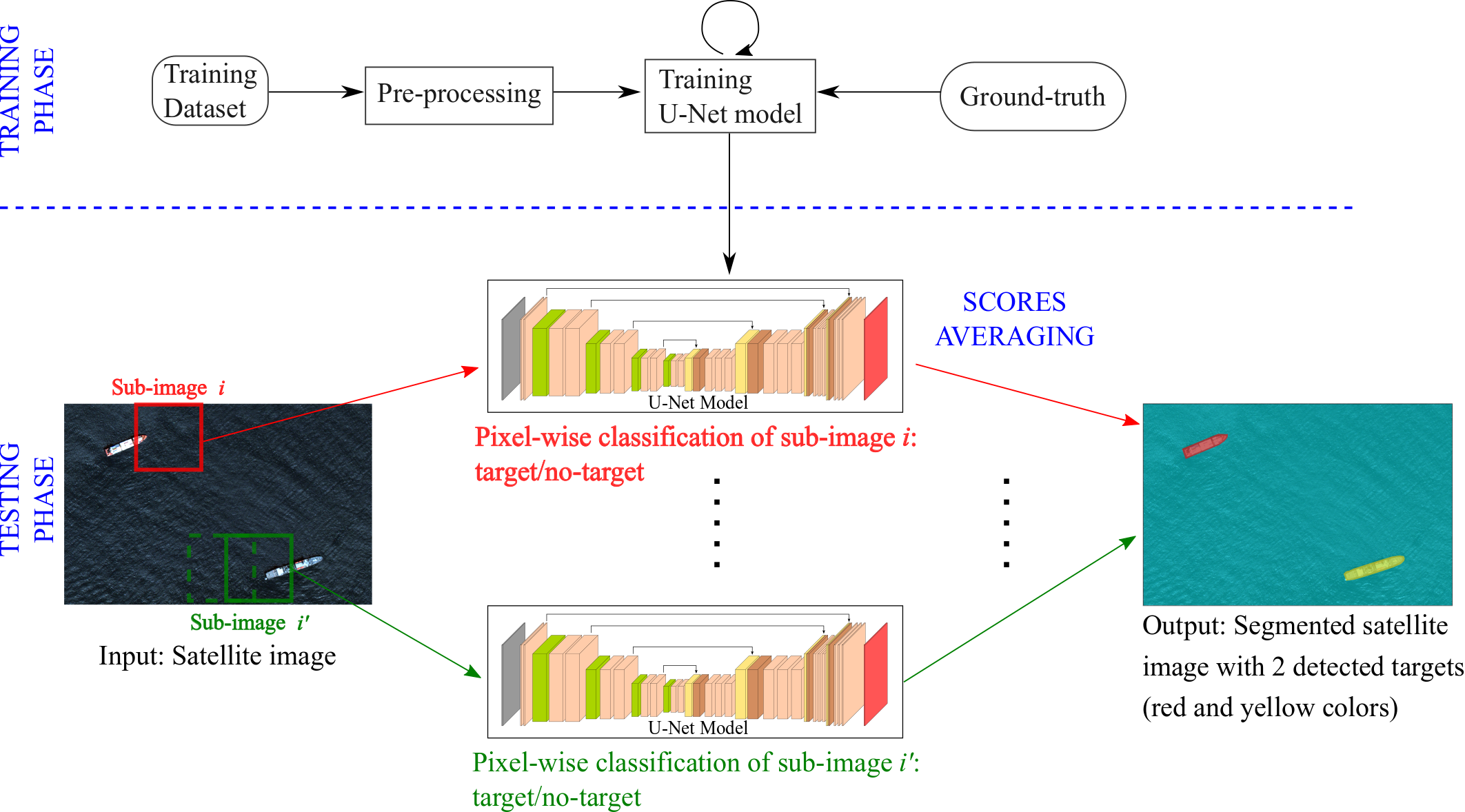}
\caption{General workflow example of a satellite image segmentation task based on the U-Net architecture. The chosen DNN architecture is trained to perform pixel-wise classification (target/no-target) using a training dataset of pre-processed satellite images that have been segmented manually in accordance with the ground-truth (training phase). During this phase, the U-Net model parameters are optimized by minimizing an appropriate cost function. The trained U-Net model is then used to perform pixel-wise classification on a test satellite image (testing phase). Concretely, a test satellite image is divided into $N$ sub-images; for each sub-image $i$, the final Softmax layer in the U-Net model assigns a classification score to each pixel. The segmented satellite image is eventually obtained by averaging the scores assigned to each pixel in the overlapping sub-images.
}
\label{fig:u-NetModel}
\vspace{-1mm}
\end{figure}

Complex DNNs with multiple hidden layers and neurons can be used for satellite image segmentation and target classification. The training of these requires large labeled datasets, and this is naturally demanding both of storage and of computational power. Graphical processing units (GPUs) are the key to address the computational demand.  
Data augmentation techniques, such as cropping, padding and flipping, enable practitioners to significantly increase the diversity of data images available to train large DNN models, without actually collecting new images~\cite[Ch. 7]{GooBenCou:B16}.
For image segmentation and target classification, convolutional neural network (CNN) architectures are the most common~\cite[Ch. 9]{GooBenCou:B16}, since they allow for an efficient analysis of image textures, i.e., the contextual signal from the neighbouring pixels of a pixel under test. During the training phase of a CNN, features automatically extracted from a set of training images are assessed to understand (learn) which of them are the most suitable and relevant to perform segmentation and/or classification. 
%The training is performed by continuously minimizing an ad-hoc cost function until a stopping criterion is satisfied, together with the application of regularization techniques for generalization of the model.
%A key feature of common CNN architectures for identifying such features are multiple and subsequent pooling operations that aggregate the feature maps derived from convolutions to a coarser spatial scale. The last layer of the network will simply contain the aggregated information whether a feature that is indicative for the target class was visible anywhere in the image or not.
Nowadays, the most used CNN architectures are the SegNet~\cite{Kendall:J15}, the U-Net~\cite{Ronnebergem:J15,CapobiancoSM18,GargiuloDIRR19}, and the Mask R-CNN~\cite{HeGkDoGi:C17}. The first two are fully-connected CNNs and they are mostly used for pixel-wise classification; the Mask R-CNN, instead, is a regional CNN mostly used to perform image segmentation by providing as output bounding boxes around the identified targets. Fig.~\ref{fig:u-NetModel} depicts a general workflow example of a satellite image segmentation task based on the U-Net architecture.

\subsection{Use Cases: AI Techniques for Satellite Image Segmentation and Classification}
\label{subsec:use_case_dataset}

\begin{figure}[!t]
\centering
\includegraphics[width=0.5\columnwidth]{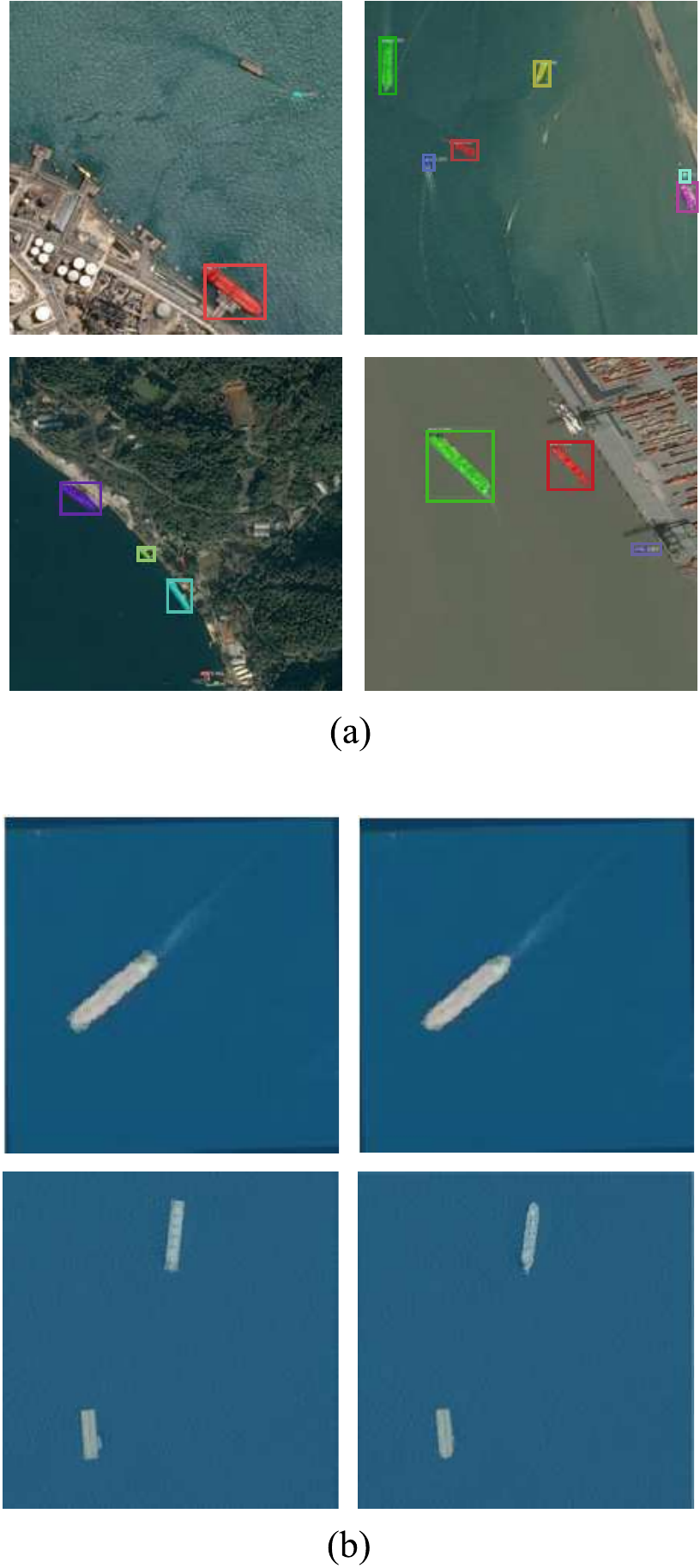}
\caption{
%\textcolor{red}{
(a) Example result of the segmentation task on VHR images (first use case): detected ships are surrounded by bounding boxes and highlighted with different colors. (b) Example result of the segmentation task ran over a validation dataset of HR images (second use case): left-hand side images represent the ground-truth, i.e., ships are manually extracted, while the right-hand side images represents the output of the segmentation task.} 
%from the synthetic dataset, on the left the ground truth and on the right the prediction are highlighted}}
\label{fig:rawSegmentation}
\end{figure}

The first use case that we present is to detect ships and extract important features, e.g, position, width, length, heading and other relevant information, from very high resolution (VHR) optical satellite images by means of CNNs. The training dataset consists of about 200000 VHR images, with spatial resolution of approximately 1.5 m and dimension 768 by 768 pixels, acquired by the GeoEye, SPOT, Pleiades and Black Sky satellites. The inference, i.e., the testing of the CNN, is performed on new images acquired by the same sensors. A Mask R-CNN architecture is used, and Fig.~\ref {fig:rawSegmentation}-(a) shows an example result of the segmentation task: detected ships are surrounded by bounding boxes and highlighted with different colors. 

The second use case is the segmentation of high resolution (HR) images, with spatial resolution of about 10 m. A real dataset of HR images for training and testing is not available. However, a synthetic dataset is obtained by degrading 10000 VHR images acquired by SPOT satellite, by means of appropriate computer vision techniques, such as the pyramid representation~\cite{adelson1984}, in which an image is subject to repeated smoothing and subsampling.  The type of architecture employed in this case is the U-Net, and Fig.~\ref{fig:rawSegmentation}-(b) represents an example result of the segmentation task ran over a validation dataset of HR images. In particular, the left-hand side images of Fig.~\ref{fig:rawSegmentation}-(b) represent the ground-truth, i.e., ships are manually extracted, while the right-hand side images represents the output of the segmentation task.

\begin{figure}[!t]
\centering
\includegraphics[width=0.5\columnwidth]{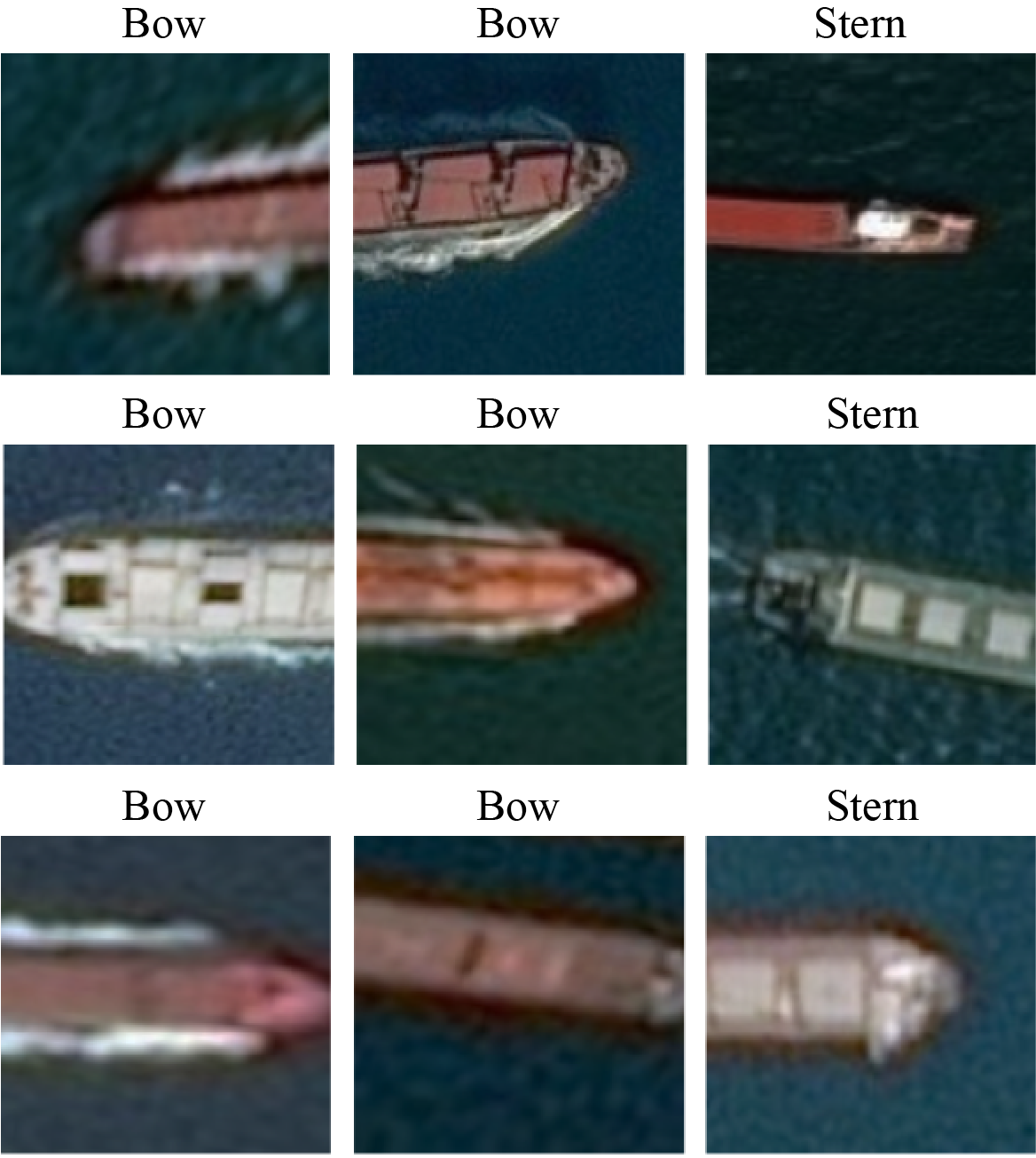}
\caption{Stern and bow classification for detected ships in clipped satellite  images acquired by S-2.}
\label{fig:SternAndBow}
\end{figure}

The third use case is related to the classification between ship stern and bow by means of a ResNet34~\cite{HeZhangRenSun:C16} architecture. The training dataset is composed of satellite images acquired by Sentinel-2 (S-2), from which detected ships are clipped, manually divided into stern and bow, and labelled. Fig.~\ref{fig:SternAndBow} represents the results of the classification.
%task performed on clipped images acquired by the S-2 satellite. %$The task is a common Image Classification with 2 labels (stern and bow) 

%\textcolor{red}{
In order to have a more complete view of the maritime situational picture, DNNs can  be also conceived to discriminate among different types of vessels, by exploiting the heterogeneous information contained in historical Sat-AIS messages. In particular, the information provided by historical Sat-AIS data, e.g., ship width, length, and type, can be used to train neural networks to classify vessels into 14 categories derived by the AIS ship type, i.e., anti pollution-law, medical-non-conflict, cargo, dredging-military-sailboat, fishing, high speed craft, other-unknown-reserved, passenger, pilot boat, pleasure, search and rescue, tanker, tug-towing, wing in ground-effect. The neural network architecture used to perform this classification task consists of 2 hidden layers, the first with 200 neurons and the second with 100 neurons, and of an output layer of 14 neurons, with each neuron providing the probability of belonging to the respective category.
During the inference phase, the neural network utilizes the information, e.g., length, width and area, provided by a segmentation task on VHR or HR satellite images, to categorize each of the detected vessels into one of the 14 classes.

\section{Extracting Knowledge from Sat-AIS Data}
\label{sec:Sat-AIS_data}

Sat-AIS is a revolutionary technology to obtain a comprehensive and global picture of maritime traffic, including that of areas far from ports and shore. 
AIS messages can be collected by low Earth orbit satellite systems and provide a global capability for maritime traffic monitoring using constellations of 
%low-cost micro- or nano-
satellites and a network of globally distributed ground-based stations.
This offers several benefits, including enhanced monitoring of fine-scale vessel behavior and traffic patterns, improved ability to identify potential threats, and more cost-effective use of assets.
The global coverage enabled by space-based technology is a key feature that led Sat-AIS to become the major source, by volume and coverage, of maritime traffic information for global-scale monitoring.
Nowadays, AIS data has been of increasing value, not only for ships themselves, but especially for coast guards, naval forces, and other marine operational authorities that can exploit such data to improve 
MS. 

The availability of global datasets provided by Sat-AIS technology opens up new ways
%As such, 
%AIS messages are an increasingly fundamental source of information 
to extract valuable knowledge for MS.
%However,  
%As a global self-reporting information source for MSA, 
%AIS data presents some challenges and limitations. 
However, with the increase of the available Sat-AIS data to massive scales, computational strategies must cope with challenges typically faced when collecting, storing, querying, and processing globally distributed datasets of considerable size. Vast amounts of data tend to overwhelm human operators and call for approaches with a high degree of automation. 
Furthermore, standard algorithms are proving to be unable to deal with the non-idealities (e.g., erroneous, incomplete, intermittent, or counterfeit information) of such datasets. 
As an example, in contrast with traditional positional measurements, Sat-AIS messages are irregularly-sampled, with an update interval usually dependent on the specific vessel's kinematic behaviour. 
Moreover, ships can enter and exit the receivers' network coverage; or, worse,
intentionally disable Sat-AIS transmitters.
All this can lead to surveillance gaps that range from several minutes to many hours. 
From this perspective, the success of future surveillance systems using maritime data will increasingly require to combine statistical, big data analytics, and AI techniques in order to handle 
all the aforementioned challenges introduced by AIS datasets.

\subsection{AIS-based Maritime Traffic Knowledge Discovery}
%\subsection{Extrapolating Statistics form Historical AIS Data}
\label{sec:ext_stat_hist_AIS_data}

\begin{figure}
\centering
\begin{minipage}{.5\columnwidth}
\centering
\includegraphics[width=.9\columnwidth]{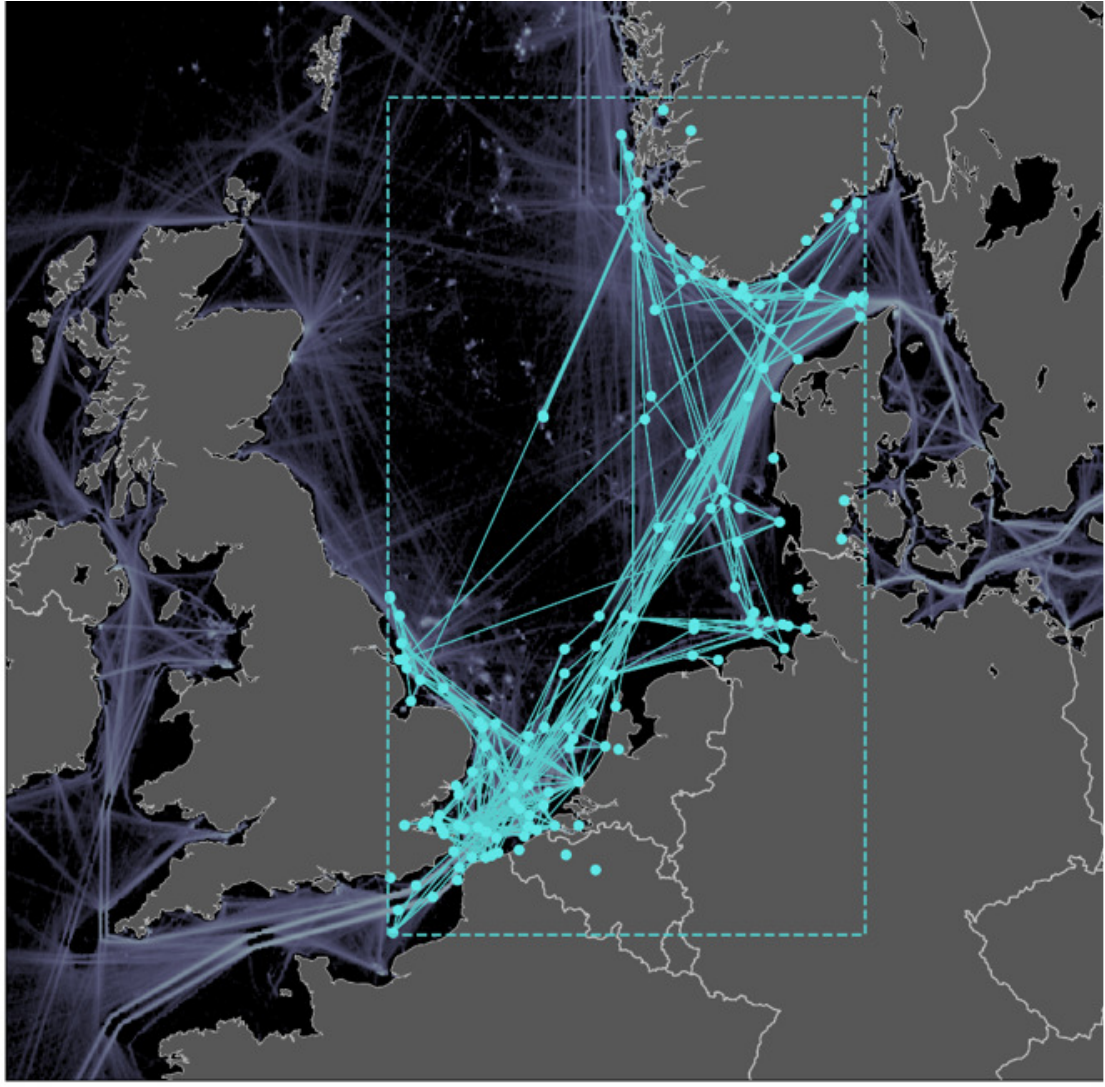}
\vspace{3mm}
\end{minipage} 
\\
\begin{minipage}{.5\columnwidth}
\centering
\hspace*{-1mm}\includegraphics[width=.9\columnwidth]{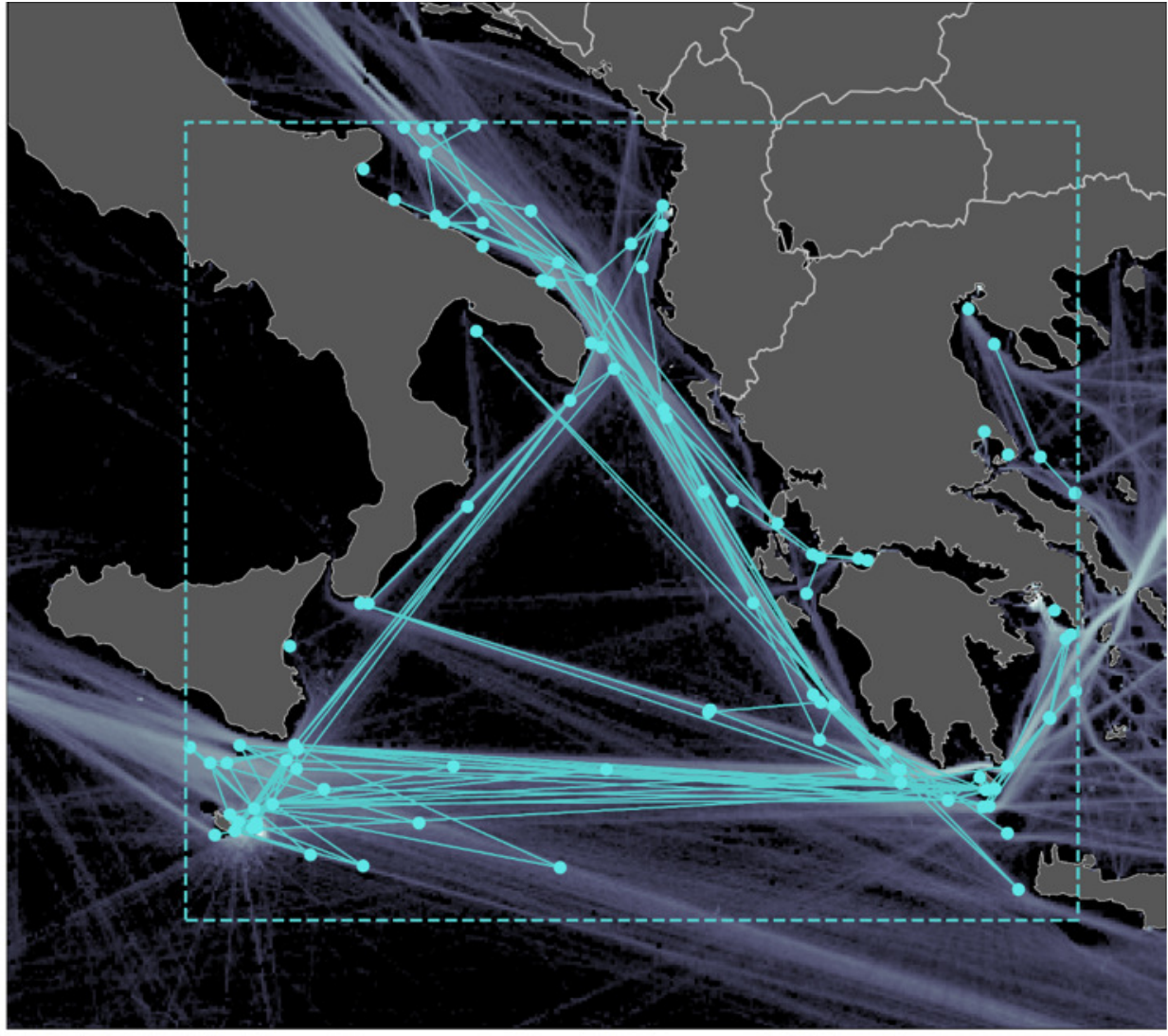}
\end{minipage}
\caption{Maritime traffic graphs (MTGs) extracted from AIS data obtained in the North sea (upper figure) and in the Mediterranean sea (lower figure).}
\end{figure}
%AIS data represents a fundamental resource for real-time monitoring of the maritime space. However, there is need of processing  the historical AIS data to extract common routes and statistics, such as destination ports, direction.

One of the biggest challenges to improve MS is capitalising on the increasingly available stream of AIS data by turning this information into 
actionable 
%intelligence and automatically extract 
knowledge of the maritime situational picture
%, providing a usable characterisation of vessels and their movements 
at global scale. 
In recent years, 
a significant body of research has been dedicated to the development of
new methodologies in support of MS
that aim at learning, from historical AIS data, motion patterns of ships, i.e., recurrent sea routes and statistical analytics about the dynamics, density and types of vessels that, if well-characterised, can be beneficial for different MS applications such as anomaly detection, knowledge-based tracking, classification, and long-term prediction of vessels.

Knowledge discovery of maritime traffic patterns from AIS data can be broadly classified \cite{survey19,Zissis20} into statistics-grid-, and vector-based methods.
In addition, machine learning tools have been recently explored to encode AIS historical (training) data for vessel trajectory prediction~\cite{Zissis2017,Nguyen2018,icassp20}.
Common statistics-based techniques \cite{Cazza16,Li16} focus on analysing the available data in order to provide a visual representation and quantitative modelling of the
fundamental statistical analytics. 
%such as traffic density, velocity and other parameters' distribution. 
Grid-based methods \cite{Bomb06,Ristic08,Xiao17}
divide the area of interest into a spatial grid of indexed cells, each dealing with the static and dynamic properties of those vessels that 
%traveled inside 
passed through
the specific element of the grid. 
This presents the benefit of reducing the overall scale of the information extraction process, as well as the storage for fast query/search operations.
%In this context, kernel density estimation (KDE) can be used to learn the distribution of kinematic variables in each cell \cite{Ristic08}.
%In addition, associative learning procedures based on biological principles have been considered 
%in \cite{Bomb06}, while a grid-based DBSCAN algorithm combined with KDE has been proposed in \cite{Xiao17}. 
%%A machine learning framework performing the tasks of clustering, classification and outlier detection has been developed in \cite{Devries} to represent vessel traffic and detect anomalous behaviours.
%Grid based methods are particularly well-suited for port security and small areas of interest due to their inherent computational limitation as the the number of grid elements increases. 
On the other hand, vector-based methods use a vectorial representation of maritime traffic, as sea routes are considered as a set of links connecting waypoints. 
This allows for high compactness of the waypoints and traffic routes representation at a global scale. 
%Vector-based methods are usually based on clustering algorithms. 
%Point clustering methods cluster the trajectory points into waypoints, connected through sea roads,
%while segment clustering techniques directly cluster trajectory segments, possibly as traffic corridors with different width, volume, and distribution \cite{Zissis20}. 
%%The historical route patterns can be associated to the navigating vessels in real time to support anomaly detection and traffic forecasting. 
A vector-based approach based on point clustering is the traffic route extraction for anomaly detection (TREAD) algorithm developed in~\cite{Pallo13}.
TREAD generates a set of historical patterns of life represented by waypoint and route features, where waypoints are defined as stationary objects like ports and offshore platforms or entry/exit points.
%A key property of TREAD is that waypoints and routes are merged, split, removed, created as new data is collected. 
%Further developments \cite{Argue18} led to a two-level representation of maritime traffic, where the inner layer provides granularity to the graph structure of the external layer. 

%Another noteworthy line of research~\cite{Coscia2018,oceans2019} recenlty developed
Successful advancements on vector-based knowledge discovery~\cite{Coscia2018,oceans2019} 
have recently led to the development of 
an unsupervised graph-based methodology 
to identify the spatiotemporal dynamics of ship routes, and
efficiently extract a compact representation of global maritime patterns
from large volumes of historical AIS data, in the form of a maritime traffic graph (MTG).
This method builds on recent advances in long-term vessel motion modeling~\cite{Millefiori2016,Millefiori2016-2} whereby the dynamics of ships can be effectively described by a piecewise Ornstein-Uhlenbeck (OU) mean-reverting stochastic process. This approach, extensively validated against real-world datasets~\cite{Millefiori2015,Millefiori2016}, relies on model parameters that change only at waypoints (places where ships regularly stop or change their velocity).
%Unlike other vector-based solutions, this method builds on recent advances in long-term vessel motion modeling \cite{Millefiori2016,Millefiori2016-2}
%that have shown, through extensive validation against real-world datasets \cite{Millefiori2015,Millefiori2016}, how the dynamics of ships along their full trajectory can be effectively described 
%by a piecewise Ornstein-Uhlenbeck (OU) mean-reverting stochastic process, whose parameters switch in correspondence of waypoints (geospatial regions where ships regularly stop or change their velocity).
The OU model makes it possible to 
statistically represent with increased accuracy the dynamics of maritime traffic,
associate sparse (in space and time) measurements to tracks,
and hence
synthesize historical ship trajectories into a sequence of waypoints
connected together by a network of
navigational legs (with non-maneuvering motion between waypoints).
%The resulting MTG, whose nodes represent waypoint areas connected together by a network of graph edges, ultimately leads to the identification and statistical characterisation of recurrent maritime traffic routes over an area of interest within a given time frame. 
The MTG method works in an unsupervised way, is computationally efficient, and can deal with big data processing models and paradigms. 
The effectiveness of the proposed methodology has been successfully demonstrated on real-world extensive AIS datasets 
collected in the Iberian Coast and English Channel areas \cite{Coscia2018},
as well as
during the validation phase in four operational trials of the EU-H2020 project for maritime integrated surveillance awareness (MARISA) \cite{oceans2019}.
Starting from raw AIS streams,
the MTG model can be automatically extracted based on the following key data processing steps as also shown in Fig.~\ref{fig:diagram_MTG}:

\begin{figure}[!t]
\centering
\includegraphics[width=.9\columnwidth]{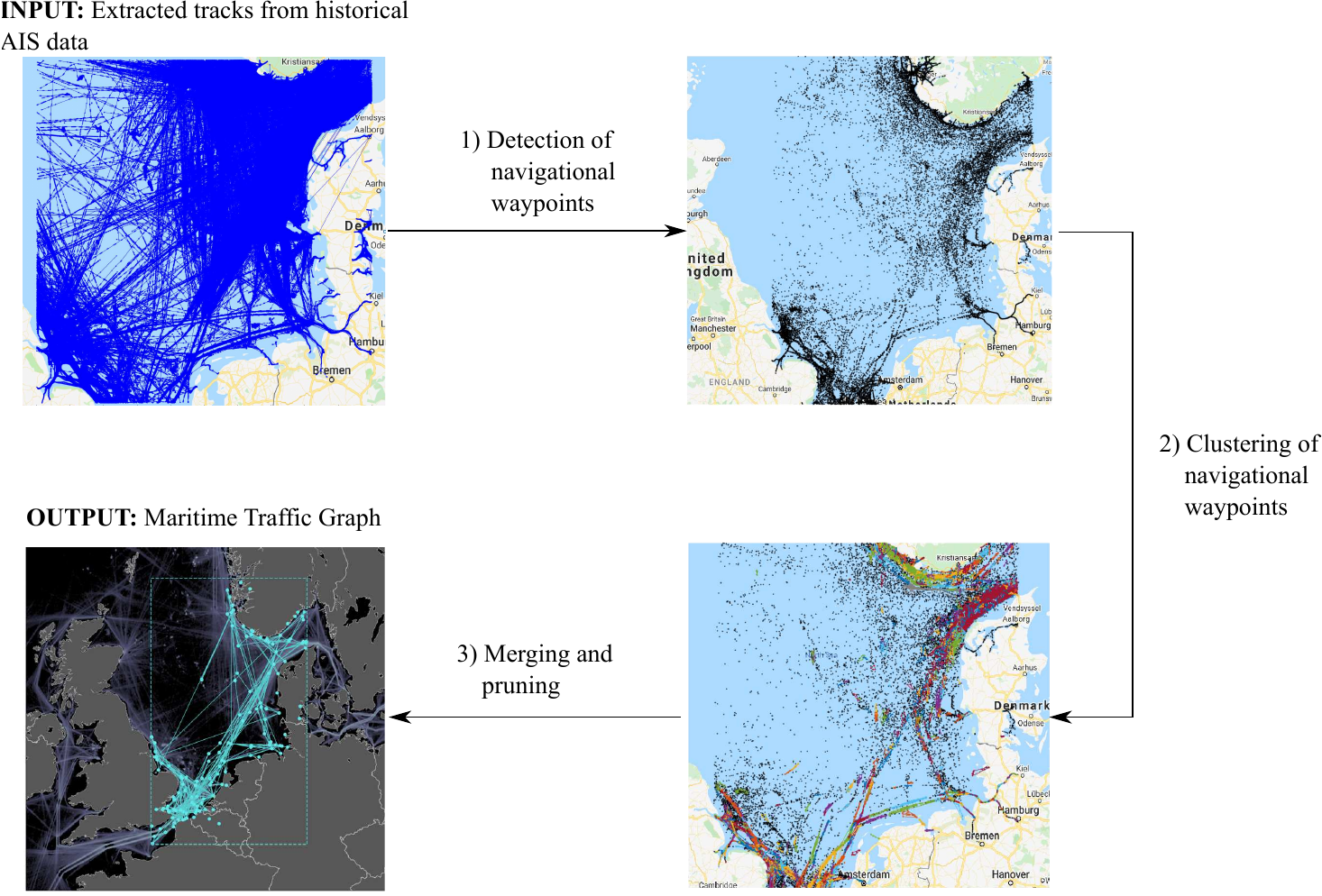}
\caption{General workflow for the extraction of a MTG model. Tracks initially extracted from historical AIS data are used as input (top-left map: 19009 AIS tracks). In the first step, waypoints for each AIS track are identified by a statistical procedure for change detection based on the OU dynamic model (top-right map: 162662 detected waypoints). In the second step, the identified waypoints are then clustered to identify specific geospatial regions where waypoints are concentrated (bottom-right map: 2286 clusters with each cluster identified by a color). During the final step pruning and merging techniques can be used to encode ships' patterns in a lower-dimensional graph representation (bottom-left map: 99 nodes and 401 edges).
}
\label{fig:diagram_MTG}
\vspace{-1mm}
\end{figure}

\subsubsection{Detection of navigational waypoints}

Based on the OU dynamic model for accurate long-term ship prediction, statistical procedures for change detection~\cite{Millefiori2017,Coscia2018} can be applied to identify specific geospatial waypoints where the mean long-term velocity parameter of the underlying OU process tends to change.
%If we assume that the long-run mean velocity of the process can abruptly change at any time instant (i.e., it is a piecewise-constant function of the time), then sequential detection procedures can be used to estimate the switching point of the long-run mean velocity 
%It is worth noting that the change detection capability depends on the specific choice of $\tau$; low values will lead to a large number of detections, whereas high values will either prevent detection or worsen the performance in terms of average run length.
%In synthesis, the procedure relies on the estimation of the long-run mean velocity parameter of the OU process, before and after a change; the CUSUM is computed in step 2) and, if a deviation from the desired threshold is observed, a change of the long-run mean velocity is declared. 
The detected waypoints represent:
i) ports, where a ship's speed is null either before (leaving the port) or after the change (entering the port);
ii) navigational waypoints, where the direction of the ship changes (while the speed is possibly constant);
iii) entry, exit, and entry/exit points, 
%which are not properly change regions, but rather 
i.e. virtual boundary regions of the area of interest that
summarise 
%take into account 
the entering and exiting traffic.
In Fig.~\ref{fig:diagram_MTG}-(top-right map) a total of 162662 waypoints are identified during this step.
%that enters and exits the given area.

\subsubsection{Clustering of navigational waypoints}
\label{sec:clustering}

Based on the assumption that most  maritime traffic is inherently regular, change points are expected to be concentrated around specific geospatial regions. 
To find these significant waypoint areas, standard density-based clustering techniques (such as density-based spatial clustering of applications with noise (DBSCAN)~\cite{dbscan}) can be used in order to group together multiple change points into a lower number of distinct waypoint clusters.
%The DBSCAN algorithm~\cite{dbscan} 
%can be adopted to associate previously detected change points to waypoint clusters based on a Mahalanobis distance that takes into account not only position, but also the direction of motion (more precisely, the angles of the long-run mean velocity vectors before and after each change).
Intuitively, a clustering algorithm will look for regions where navigational change points are very close in the feature space and will identify outliers as points lying in low-density regions.
In Fig.~\ref{fig:diagram_MTG}-(bottom-right map) a total of 2286 clusters are found during this step (each cluster is identified by a color).
%A large volume of historical AIS data is usually necessary to generate the traffic graph. 
%However, as the amount of available AIS data grows to massive scales, computational techniques are needed to efficiently process, manage and store data.
%In this regard, we developed a fast and scalable implementation of DBSCAN that computes only distances between change points that are \textit{nearest neighbors}, i.e., such that the distance metric between those data points is less than or equal to a predefined radius. This allows us to avoid computing the full large-scale matrix of all-to-all distances.

%\subsubsection{Extraction of maritime patterns}
%
%Starting from a collection of raw AIS messages, a ship trajectory can be represented
%by means of a set of waypoints and a piecewise constant profile for the long-run mean velocity.
%This allows us to extract a low-dimensional representation of maritime traffic
%that can be encoded through
%a directed weighted graph, whose nodes represent navigational waypoints, while edges represent navigational legs.
%Each edge is assigned a weight, proportional to the number of vessels which transitioned from the source to the target node of that specific link.
%By using this graph-based model, we can automatically extract maritime traffic patterns from historical AIS data.

\subsubsection{Merging and pruning procedures}

When clustering algorithms are applied to large-scale real-world datasets involving a huge number of data points,
the output generated by such unsupervised classification will usually need some post-processing.
To this end, pruning and merging techniques can be used to progressively improve and simplify the overall MTG by reducing the number of graph entities (i.e., nodes and edges), and thus implicitly encode knowledge about ships' patterns using a lower-dimensional representation.
The number of graph edges can be reduced by eliminating those links characterized by low weights, these being least likely to represent recurrent patterns; and by merging closely-spaced
edges (connecting waypoints in a cluster with waypoints in a different cluster) into
one, as these are more efficiently represented by a single route.
Fig.~\ref{fig:diagram_MTG}-(bottom-left image) represents the resulting MTG which consists of 99 nodes and 401 edges.
%Moreover, edges falling over land are removed by a land avoidance logic based on bathymetry data
%and statistically close clusters (i.e., that fall within a given Mahalanobis distance) can be grouped together into a single waypoint node.

%\subsubsection{Graph structure and attributes}

%Each edge is assigned a weight, proportional to the number of vessels which transitioned from the source to the target node of that specific link.
As a result, the structure of maritime traffic in the area of interest during a reference time interval 
can be represented, using graph formalism, 
by a directed graph
$\mathcal{G} = (\mathcal{N},\mathcal{E})$
where $\mathcal{N} = \{1,2,...,N\}$ is the set of nodes, and $\mathcal{E} \subseteq \mathcal{N} \times \mathcal{N}$ is the set of edges.
In particular, it is supposed that $(i, j)$ belongs to $\mathcal{E}$ if and only if there are ship tracks with two consecutive waypoints $i$ and $j$, where node $i$ is the predecessor of node $j$.
In this way, the adjacency matrix of the graph
$\mathcal{A}$
can be directly constructed from the raw ship tracks data (i.e., time-ordered lists of AIS messages) by simply identifying the transitions (and associated direction) of ships from a generic pair of nodes.
We naturally assume that $(i,i) \notin \mathcal{E}$ for any $i \in \mathcal{N}$, so that diagonal elements
of $\mathcal{A}$  are set to zero, i.e. $\mathcal{G}$ is a directed simple graph
with no self-loops.
A graph of order (i.e., number of nodes) $|\mathcal{N}|=N$ and size (i.e., number of edges) $|\mathcal{E}|=L$ is denoted as $\mathcal{G}(N,L)$.
In conclusion, the extracted waypoints -- nodes and edges of the MTG -- can be represented as vector features with different attributes directly extrapolated from the available AIS messages or computed during the graph extraction phase. 
%The attributes of each node entity may include 
%the waypoint identifier and type, its geographical location, the traffic data category, analytics about the inward/outward speed of ships, the volume of traffic
%passing through the waypoint, etc.
The attributes of each graph entity will include aggregate georeferenced data and statistical analytics about the identified maritime patterns. 
The resulting geospatial information layer, that can be directly used for efficient mapping, query and search operations, serves as a baseline reference of maritime traffic for different MS applications.

\subsection{Maritime Anomaly Detection}
\label{sec:anomaly_detection}

%The Ornstein-Uhlenbeck (OU) mean-reverting stochastic process 
%described in Section~\ref{sec:OU_model} 
%can be also exploited 
%to detect illegal activities, for example, smuggling, drug exchange or Illegal, Unreported and Unregulated (IUU) fishing. 
%For instance, a vessel deviates from a planned route, changing its nominal velocity, stops in the middle of the sea to exchange drugs, and then restarts getting back to its nominal velocity and route. In order to hide this behavior, the vessel switches off its AIS device for a certain time, until he gets to the normal behavior. 
%The main challenge consists of  deciding if a deviation happened or not, relying only upon two consecutive AIS contacts or two radar contacts, or two consecutive SAR images. 
%The OU model allows to build a statistical framework in which a proper hypothesis testing procedure based on the changes in the OU process long-term velocity parameter of the vessel enables to detect the illegal activity.

Contextual information about historical maritime traffic extracted through knowledge discovery methods discussed in Section~\ref{sec:ext_stat_hist_AIS_data} 
can be highly beneficial
as prior information on the nominal dynamic behaviour of vessels
to determine when anomalous activities are happening.
Common threats in the maritime domain include
drug smuggling, piracy and terrorism, 
illegal
immigration,
marine pollution, 
prohibited imports/exports, or illegal, unreported and unregulated (IUU) fishing.
Recently, anomaly detection strategies have been explored and applied in maritime traffic monitoring
\cite{Ristic08,Lane10,Kowalska12,Vespe12,Enrica18,Enrica18-B}
in order to detect unexpected ship stops or unexpected changes in course, i.e. any vessel's anomalous deviation from the standard route
that might be related to an activity that requires closer attention.
Most research on maritime anomaly detection~\cite{Ristic08,Lane10,Kowalska12,Vespe12}
relies on two steps: i) knowledge discovery of maritime traffic patterns from historical data, and
ii) anomaly detection via unsupervised learning. 

More recently, an innovative statistical framework \cite{Enrica18,Enrica18-B} that combines the available
context data with a parametric model of the vessel's kinematic behaviour has received special attention. 
Unlike other works \cite{Ristic08,Lane10,Kowalska12,Vespe12}, here the formulation of maritime anomaly detection is based on the OU dynamic model \cite{Millefiori2016}. 
While it is useful that the OU model represents targets' motion in terms of speed and direction, its key strength 
is its greatly reduced prediction uncertainties: deviations from prediction are less easily explained as random 
noise than they would be if other models were used.
%In this respect, a first line of research 
This line of research 
has focused on 
%detecting anomalous deviations from standard navigational routes, even if vessels shut off their AIS transponder. 
the development of anomaly detection strategies that provide the ability to 
reveal vessel deviations from standard navigational routes as well as \textit{dark} deviations in the presence of possibly intentional
surveillance gaps (due to AIS disablements or limited sensor coverage). 
%with given probabilities of detection and false alarm.
Using the OU model to represent the vessel dynamics and contextual information to define the nominal behaviour at sea, the anomaly detector 
in~\cite{Enrica18,Enrica18-B} 
runs a hypothesis testing procedure -- the generalized likelihood ratio test --
to make decisions on the existence of anomalous deviations within a given time window relying upon the available measurements (e.g. AIS, radar, SAR).
This method, successfully applied to a real-world rendezvous detection scenario in \cite{Enrica18-B}, can integrate measurements from multiple sources and handle different levels of data unavailability.
The use of multiple heterogeneous measurements associated with the vessel under track, e.g., AIS messages combined with non-cooperative contacts from coastal surveillance radar networks or satellite networks,
can lead to a significant improvement of detection performance.
From the end-user perspective, this method enables operators to establish the detection false alarm rate, which is a well-known issue for maritime command and control systems.
Moreover, such an automatic detector enables quick reaction of human operators to anomalies at sea with much improved chances of a productive operation for surveillance and intercept units.

%The second line of effort focused on the extension of 
Further advances led to the design of 
%filtering versions~\cite{fusion18} for  
a probabilistic joint anomaly detection and tracking methodology \cite{fusion18,icassp19}
for sequential detection of maritime anomalous deviations and simultaneous vessel tracking
where new contacts are periodically available based on surveillance coverage and reporting frequencies. 
This is motivated by the fact that maritime anomaly detection would ideally be performed in real-time, as new observations become available.
%This data possibly includes spurious measurements of different nature.
This method allows for a mathematical definition of nominal vessel behaviour based on the combination of the OU dynamic model and context information extracted from historical data.
The key idea behind this approach is to represent any anomalous deviation as a switching unknown control input that goes into action by modifying the nominal object dynamics, 
i.e., the OU mean velocity parameter.
As a result, anomaly detection and target tracking can be recast as a
special Bayesian state estimation problem for which recent advances on hybrid Bernoulli filtering~\cite{rnc2019} can be used.
%introduced for systems with unknown switching (on/off) inputs.
This Bayesian framework has been extended in \cite{icassp19} 
where an adaptive version,
based on the multiple-model approach \cite{BarWilTia:B11}, 
of the single-model 
joint anomaly detection and tracking filter
has been derived 
to handle unknown parameters in the underlying vessel dynamics.

\section{Bayesian Information Fusion and MTT for MS}
\label{sec:radars_ms_overview_info}

\begin{figure}[!t]
\centering
\includegraphics[width=.5\columnwidth]{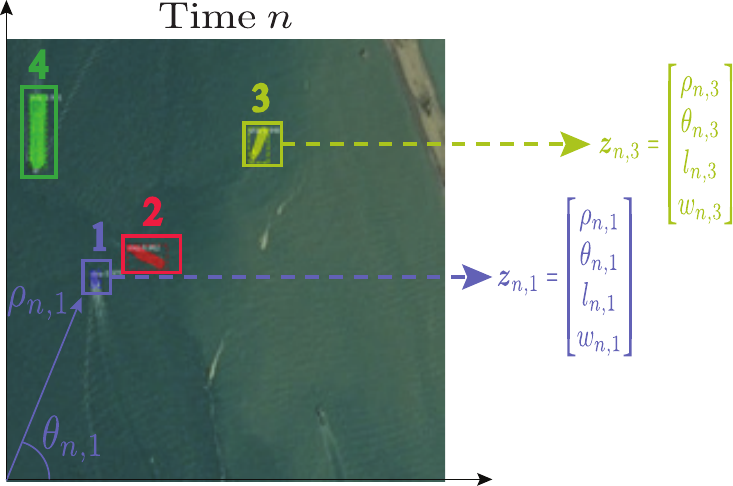}
\caption{Example of measurements $\V{z}_{n,m}$ (sensor index $s$ is omitted for clarity) extracted from four detected ships in a MSP image acquired by S-2 satellite. In this example scenario, each measurement $\V{z}_{n,m}$ consists of the range $\rho_{n,m}$, the bearing $\theta_{n,m}$, length $l_{n,m}$ and the width $w_{n,m}$ of the detected ships.}
\label{fig:example_measurements}
\end{figure}

%The increasing availability of space-based remote sensors providing continuous coverage on remote areas of the Earth requires the development of sophisticated information fusion and MTT algorithms, which are able to combine and fuse information from different heterogeneous sources, e.g., terrestrial radars, Sat-AIS, SAR, and optical sensors.
%% For the development of future MSA systems that combine multiple HFSW radars with other sources of information, there is a need for dedicated 
%% multitarget tracking and information fusion algorithms. 
The main purpose of a multisensor MTT algorithm is to sequentially determine the number of ships in the MS area and estimate their states, e.g., position, velocity, course, and heading, by exploiting  measurements from multiple heterogeneous sensors.  The measurements are noisy observations of the kinematics, dimensions, shapes, or other features of the targets that can also be output of a classifier. The MTT problem thus consists of a detection step and an estimation step; mathematically, these can be formalized as follows. We denote with $\V{x}_n = [\V{x}_{n,1}^\T \cdots \V{x}_{n,K_n}^\T]^\T$ the unknown states of $K_n$ targets at time $n$, with $\V{z}_n^{(s)} = [\V{z}_{n,1}^{(s)\T} \cdots \V{z}^{(s)\T}_{n,M_n^{(s)}}]^{\T}$  the $M_n^{(s)}$ measurements generated by sensor $s$ at time $n$ and with $\V{z}$ all the measurements from all sensors up to time $n$. 
Fig.~\ref{fig:example_measurements} represents examples of measurements $\V{z}_{n,m}^{(s)}$ that could be extracted from a MSP image acquired by a S-2 optical sensor by means of a segmentation and classification technique, as described in Section~\ref{subsec:main_architectures_obj_class}.  
Based on the measurements $\V{z}$, the likelihood functions $f(\V{z}_{n,m}^{(s)} | \V{x}_{n,k})$ and the probability density functions (pdfs) $f(\V{x}_{n,k} | \V{x}_{n-1,k})$, e.g., the OU dynamic model, we wish to
\begin{enumerate}
\item \textit{(Detection step)} estimate the number of targets $\hat{K}_n$;
\item \textit{(Estimation step)} estimate the target states $\hat{\V{x}}_{n} = [\hat{\V{x}}_{n,1}, \ldots, \hat{\V{x}}_{n,\hat{K}_n}]$.
\end{enumerate}
In a Bayesian formulation, the estimation step 
%% target detection and state nestimation step 
essentially amounts to calculating at each time $n$,
%the posterior existence probabilities $p(r_{k} \!=\! 1|\V{z})$ and 
the posterior pdfs $f(\V{x}_{n,k} | \V{z} )$ of the states $\V{x}_{n,k}$ given all the measurements up to current time $\V{z}$. 
MTT methods have to cope with various challenges, for example, the heterogeneity of the different information sources \cite{VivoneBH15_53,MeyBraWilHla:J17,MeyKroWilLauHlaBraWin:J18},
and the measurement-origin uncertainty (MOU), i.e., the fact that it is unknown which target (if any) generated which measurement.
%, and, the methods have to be flexible enough to exploit, if available, additional available information, e.g., target dimension, shape, classifier output, such as the  probability indicating whether a target belongs to a certain category among a set of predefined categories.
Existing MTT algorithms can be broadly classified as ``vector-type'' algorithms, such as the joint probabilistic data association (JPDA) filter \cite{BarWilTia:B11} and the multiple hypothesis tracker (MHT) \cite{Rei:J79,ChongMR18}, and ``set-type'' algorithms, such as the (cardinalized) probability hypothesis density 
%%H (PHD) 
filter \cite{Mah:B07,Mah:J03,Mah:J07,VoVoCan:J07}
%%H probability hypothesis density (PHD) filter \cite{Mah:B07}, the cardinalized PHD (CPHD) filter \cite{Mah:B07}, 
and multi-Bernoulli filters \cite{Mah:B07,VoVoCan:J09}.
%% Wil:J15}. 
Vector-type algorithms represent the multitarget states and measurements by random vectors, whereas set-type algorithms represent them
%% the multitarget states and measurements 
by random finite sets.
%%H  (RFSs). 
Algorithms of both types have been developed and evaluated~\cite{PonsforW10_18,MarescaBHG14_52}.
%%% However, these algorithms
Several limitations have been noted. First, the fusion of heterogeneous
information sources is not straightforward. Second, they do not adapt to time-varying model parameters. And third, their complexity usually does not scale well in relevant system parameters, e.g., the number of sensors.
%% , and thus they are impractical for a large number of sensors and/or targets.  
%%  Extending these algorithms such that they can incorporate data from an additional source or can adapt to time-varying environmental conditions, would essentially mean 
%% to derive a new algorithm from scratch. 

\subsection{SPA-based Multisensor MTT Algorithm} %% \\ for Multiple Dynamic Models}
\label{sec:multisensor_multitarget_tracking_SPA}
An emerging approach to MTT and information fusion -- one with flexibility, low complexity and useful scalability -- is based on a factor graph and the \emph{sum-product algorithm} (SPA)~\cite{MeyBraWilHla:J17,MeyKroWilLauHlaBraWin:J18,GagSolMeyHlaBraFarWin:J20}.  
First, a factor graph representing the statistical model of the MTT problem is derived; then, the SPA is used to obtain a principled and intuitive approximation of the Bayesian inference needed for targets detection and estimation.
A major advantage of the SPA is its ability to 
%%% systematically
exploit conditional independence properties of random variables for a drastic reduction of complexity; thereby, SPA-based MTT algorithms can achieve an attractive 
performance-complexity compromise (see \cite{MeyKroWilLauHlaBraWin:J18} and references therein), making them suitable for large-scale tracking scenarios involving a large number of targets, sensors, and measurements, and allowing their use on resource-limited devices.
%%%
The SPA's  versatility and intuitiveness 
%%% of the SPA methodology 
%% make it possible to develop 
has enabled the establishment of a suite of Bayesian multisensor MTT tracking and information fusion algorithms 
where, 
%%% somewhat 
similar to 
%%% the elements of 
a construction kit system, 
%%% existing 
algorithm parts can be combined, extended, or adapted to achieve desired functionalities and properties. 
SPA-based MTT algorithms can be extended to fuse heterogeneous data, e.g., terrestrial radars, SAR, optical sensors, and AIS. They can incorporate different dynamic models such as nearly constant velocity (NCV) or OU. They can also be designed to infer time-varying model parameters, such as detection probabilities of radar sensors; and to select an index from a menu of multiple target motion models.
The use of SPA-based MTT algorithm is promising due to the highly efficient solution of the MOU problem combined with sequential Monte Carlo techniques, and is potentially suitable for arbitrary non-linear and non-Gaussian problems.
%% The factor graph provides a graphical representation in which different functional units of an
%% algorithm appear as distinct parts. This introduces a desirable lucidity and modularity into algorithm design.
%This section provides an overview on SPA-based multisensor MTT algorithm highlighting its flexibility to adapt to use measurements and information from various heterogeneous sources.
%%% introduced in 
%%% \cite{MeyBraWilHla:C15,
%\cite{MeyBraWilHla:J17, MeyKroWilLauHlaBraWin:J18, soldi18}.
The SPA-based MTT algorithm estimates the number of targets $K_n$ and their state at each time $n$ by efficiently  solving the MOU problem and calculating the beliefs $\tilde{f}(\V{x}_{n,k})$, which approximates the marginal posterior pdfs $f(\V{x}_{n,k} | \V{z})$, by employing an iterative version of the SPA on a suitably devised factor graph~\cite{MeyBraWilHla:J17,MeyKroWilLauHlaBraWin:J18}.%similar to the one depicted in Fig.~\ref{fig:exampleDistribution}. 

\begin{figure*}[!t]
%\centering\hspace{4.5mm}
\centering

\includegraphics[width=0.95\columnwidth]{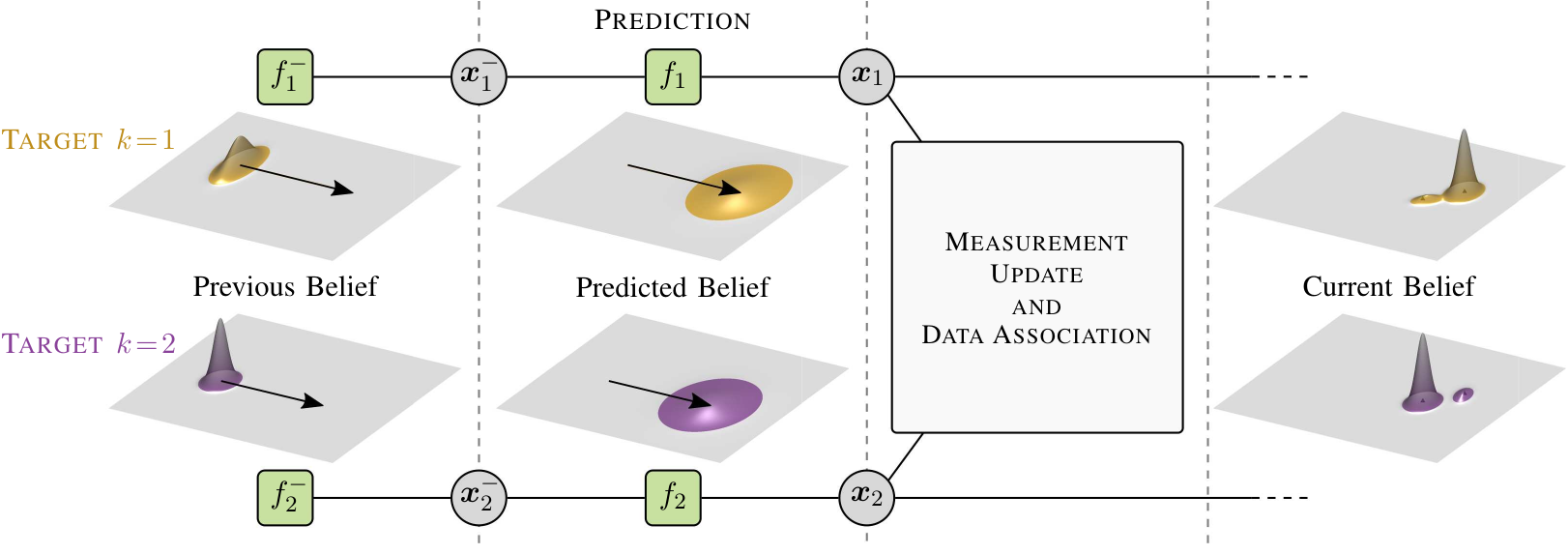}
%\end{postscript}
\caption{An example of factor graph 
for the case of two targets and two measurements produced by a single sensor.
%% each originating from a corresponding target.
In addition, the previous, predicted, and current beliefs for the two targets are shown.}
\label{fig:exampleDistribution}
\end{figure*}

The SPA approach is illustrated in Fig.~\ref{fig:exampleDistribution} via a simple 2-target example with two measurements and a single sensor.
%% each originating from a corresponding target.
%% Fig.\ \ref{fig:exampleDistribution} depicts
%The factor graph representing the factorization~\eqref{eq:factorization} for
%%% the case of two targets and two measurements produced by a single sensor, each originating from a corresponding target.
%this case is shown in Fig.~\ref{fig:exampleDistribution}.
%% The SPA for one time step $n$ consists of a \emph{prediction step} and a \emph{measurement update and data association step}, which correspond to two different stages of the factor graph, and it converts the previous belief (approximating $f(\V{y}_{k,n-1} | \V{z}_{1:n-1})$) into the current belief  (approximating $f(\V{y}_{k,n} | \V{z}_{1:n})$).
%% The factor graph 
It is structured into four sections that refer to various beliefs and/or operations within the SPA-based MTT  algorithm.
%% \begin{picture}(\textwidth,100mm)
%% \begin{wrapfigure}{l}{1.1\textwidth}
The \textit{previous belief} for target $k \! \in \! \{1,2 \}$ approximates the marginal posterior pdf of the previous state $\V{x}_{k,n-1}$, i.e., $f(\V{x}_{k,n-1} | \V{z}^-)$, 
%% $k \! \in \! \{1,2 \}$, 
which is represented by the factor node ``$f^{-}_{k}$'' in Fig.~\ref{fig:exampleDistribution}, and where $\V{z}^-$ denotes all the measurements up to time $n-1$.
%% is represented in the factor graph by the factor node ``$f^{-}_{k}$''. 
In the \textit{prediction} step, the previous belief is converted into the \textit{predicted belief}, which approximates the pdf $f(\V{x}_{n,k}|\V{z}^-)$. The prediction is performed by utilizing an appropriate dynamic model $f(\V{x}_{n,k}|\V{x}_{n-1,k})$, which is represented by the factor node ``$f_{k}$''. When the time interval between two consecutive time steps is large, e.g., in the order of hours, as for the case of two consecutive SAR images, the OU dynamic model can be used to reduce the prediction uncertainty.
In the \textit{measurement update and data association} step, the measurements $\V{z}_{n,1}^{(1)}$ and $\V{z}_{n,2}^{(1)}$ are used to evaluate the likelihoods $f(\V{z}_{n,m}^{(1)}|\V{x}_{n,k})$ and solve the MOU problem as described in~\cite{MeyKroWilLauHlaBraWin:J18}. %by using the SPADA algorithm.
This results in the \textit{current belief}, which approximates the current marginal posterior pdf $f (\V{x}_{n,k} | \V{z})$ at time step $n$. 
%%% , i.e. $b(\V{y}_{1,n}) \approx f (\V{y}_{1,n} | \V{z}_{1:n})$ and $b(\V{y}_{2,n}) \approx f (\V{y}_{2,n} | \V{z}_{1:n})$. 
The current belief is centered around the current measurement and it is used for target detection and 
for state estimation at the current time step $n$. 
In addition to the factor graph, Fig.\ \ref{fig:exampleDistribution} also visualizes the previous, predicted, and current beliefs for the first and second target in the upper and lower three-dimensional plots, 
respectively. The arrows in the lefthand plots represent the trajectories of the targets.
Because of the uncertainty of target-measurement association and the proximity of the two targets, 
the current beliefs are bimodal.
The smaller of the two modes is centered roughly at the position of the respective other target.

\subsubsection{Multiple Dynamic Models and Integration of Contextual Information}
\label{subsubsec:adapting_dyn_models}

Many tracking scenarios, such as those involving maneuvering targets, require the use of different dynamic models in different time periods in order to better describe targets that maneuver, such as alternating between NCV and constant turn-rate motion models.
Therefore, following an interacting multiple model approach \cite[Ch.~11]{BarRonKir:01},  the evolution of the state of a target can be modelled by means of a set %
%${\{ \mathcal{D}_{j} \}}_{j=1}^J$ 
of possible dynamic models. A dynamic model (DM) index $\ell_{n,k}$ is introduced for each target $k$  to select the most appropriate DM at time $n$~\cite{SoldiB:C18,SoldiMBH:J19}. %$\mathcal{D}_{j}$.
%\vspace{-1mm}
%as
%\begin{equation*}
%\V{x}_{k} =\ist \bm{\xi}_{\ell_{k}}\rmv\big(\V{x}_{k}^- \rmv,\V{u}_{k}^{(\ell_{k})}\big)\ist.
%%% \label{eq:dyn_models}
%\vspace{-1mm}
%\end{equation*}
%Here, $\bm{\xi}_{\ell_{k}}(\ist\cdot\ist,\cdot\ist)$ is the state-transition function of PT $k$ that is in force at the current time and $\V{u}_{k}^{(\ell_{k})}$ 
%is a driving process that is assumed independent and identically distributed (iid) across time and 
%%% PT 
%$k$ \cite{BarWilTia:B11,BarRonKir:01}.
%The DM $\mathcal{D}_{\ell_{k}}$ is then defined by $\bm{\xi}_{\ell_{k}}(\ist\cdot\ist,\cdot\ist)$ and the statistics of $\V{u}_{k}^{(\ell_{k})}\rmv$, 
%and it is selected from a set ${\{ \mathcal{D}_{j} \}}_{j=1}^J$ of possible DMs by 
%% From $\mathcal{D}_{j}$, one can obtain the state-transition pdf $f_{j}(\V{x}_{k}|\V{x}_{k}^-)$. ???DON'T WE NEED $f_{j}(\V{y}_{k}|\V{y}_{k}^-)$???
%From $\mathcal{D}_{j}$, one can obtain the state-transition pdf $f_{j}(\V{x}_{k}|\V{x}_{k}^-)$.
%%%  and
%%% , consequently, the augmented state-transition pdf 
%%% $f_{j}(\V{y}_{k}|\V{y}_{k}^-)$. 
The DM indices $\ell_{n,k}$
are modeled as discrete random variables that are independent across targets and evolve in time 
%% temporally 
according to a Markov chain.
%The use of multiple DMs and the introduction of a DM index $\ell_{n,k}$ allows to better track targets even in cases where they perform sharp maneuvering. 
Bayesian inference on the DM indices $\ell_{n,k}$ and the target states $\V{x}_{n,k}$ can be still be performed by running the SPA algorithm on a suitable devised factor graph.

Similarly, the multiple DM formalism can be used to improve the performance of an MTT algorithm by integrating geographic information about standard maritime routes, e.g., MTGs obtained by historical Sat-AIS data, as explained in Section~\ref{sec:ext_stat_hist_AIS_data}. 
%% We will show how this 
%This type of information fusion can be accomplished by using the ``multiple DM'' formalism of Section \ref{subsec:adapting_dyn_models} with suitable modifications. 
In particular, each navigational leg of an MTG, i.e., an edge $(i,j)$ connecting waypoint $i$ to waypoint $j$, can be associated to a specific dynamic model characterized by a dominant direction, that is, from waypoint $i$ to waypoint $j$. Still, a discrete random variable $\ell_{n,k}$ can be introduced  to select the DM, that is the maritime route, that the target is following at time $n$~\cite{SoldiGMHBFW:C19}.
The evolution of $\ell_{n,k}$ is again model
ed by a Markov chain.  
%However, because the number of target routes $J_\text{R}$ can be very large, we use a reduced Markov chain 
%corresponding to a \textit{variable structure interacting multiple model} 
%%% (VS-IMM) 
%\cite{VivoneBH15_53}. The idea is to consider
%%% , at each time, 
%index transitions only to those 
%%% of the 
%target routes that 
%%% (according to some criterion) 
%are currently close enough so that they can be reached by the respective PT $k$. 
%The set of these target route indices $j$ depends on the PT index $k$ and the previous PT state 
%$\V{x}_{k}^-\rmv$, and thus also the transition matrix $\V{L}_k(\V{x}^-_k)$ depends on $k$ and on $\V{x}_k^-$. 
This results in a multisensor MTT algorithm that 
automatically takes into account the knowledge of standard
%% available 
maritime routes. 
%The fact that only a subset of all the DM index transitions is allowed
%typically leads to a significant reduction of complexity and also an improved tracking accuracy. 
%A difference from Section \ref{subsec:adapting_dyn_models}
%is the fact that the possible index transitions and the corresponding transition matrix $\V{L}_k(\V{x}^-_k)$ have to be determined at each time step and for each PT $k$. 

\subsubsection{Fusion of Sat-AIS information}
\label{subsec:surveillance}

%% In this section we show how the SPA-based MTT algorithm can be extended to accommodate heterogeneous data sources and integrate geographic information.

The  SPA-based MTT algorithm can also be extended to fuse information coming from a Sat-AIS system~\cite{GaglioneBS:C18,SoldiGMHBFW:C19}. The fusion of this information is often difficult due to the asynchronicity and sparsity of the Sat-AIS messages, and the non-trivial association between messages and targets. Indeed, although each Sat-AIS message usually includes a unique maritime mobile service identity (MMSI), this may be absent, or mistaken for a different MMSI, or observed for the first time, in which case no prior information is available on the target-message association. 
The SPA-based MTT method can be efficiently extended to fuse Sat-AIS messages and measurements obtained from SAR, optical images, or other sensors, and to identify (or label) each detected ship by means of the MMSI. The label associated to each target is modelled as a discrete random variable: marginal posterior pdfs of the targets states $\V{x}_{n,k}$ and of the labels can still be efficiently obtained by running the SPA algorithm on a factor graph derived by the underlying statistical formulation of the problem.

\subsubsection{Classification-Aided SPA-based Multitarget Tracking}
\label{subsec:class_aided_multitarget_tracking}

As explained in Section~\ref{subsec:main_architectures_obj_class}, modern deep learning segmentation and classification techniques are able to provide, in addition to kinematic data, class information for detected ships in satellite images. This information assigns each ship to one category among a predefined finite set of categories, e.g., cargo, dredging-military-sailboat, fishing, high speed craft, other-unknown-reserved, passenger or pilot boat. %In the maritime domain, for example, such categories could be commercial ships, military ships, or fishing boats.
The SPA-based MTT approach can be efficiently extended for the exploitation of imperfect
\textit{class} information. 
%\cite{Drummond:C01,BarShalomKirGok:J05_41}, which assigns a target 
%to one category among a predefined finite 
%set of categories. 
This class information allows, for instance, the use of 
class-dependent DMs or measurement models. As a consequence, the inclusion of class information greatly improves the performance of the SPA-based MTT algorithm~\cite{GagSolBraMagMeyHla:J19,SoldiGDBSFTP:C20}.
%Classification-aided tracking techniques \cite{Drummond:C01,BarShalomKirGok:J05_41,RavindraBD09,SinghTDPW09_39,YingZS11_1,GeorgescuW13,
%Mellema14,MoriCC14_50} have lately encountered a growing interest, in particular following recent advances in deep learning classification methods \cite{HuangZDZ:J12,ChenLZWG:J14,LeCunBH:J15}.
%In~\cite{SoldiICassp:2019}
%% is to 
%we have shown how imperfect
%% the 
%target class information can be efficiently integrated into the 
%% recently proposed 
%SPA-based MTT method of \cite{MeyBraWilHla:J17,MeyKroWilLauHlaBraWin:J18} and improve drastically the performance of an SPA-based MTT algorithm~\cite{MeyBraWilHla:J17}. 
%% enhancing the target-measurement association and improving the overall tracking capabilities.
%% and to demonstrate the resulting performance improvement.
%% Experimental results obtained in a simulated scenario will show how the 
%More concretely, we have proposed an SPA-based 
%%% framework can efficiently employ 
%MTT method that takes into account the output of a
%%% target/clutter measurement
%classifier
%%% which 
%distinguishing between target-related classes as well as between target- and clutter-originated measurements.
%% to improve the 
%% data association and discriminate between true target tracks and false target tracks, originated by clutter measurements, 
%% target tracking accuracy 
%The proposed classification-aided MTT method outperforms the 
%%% baseline SPA-based 
%MTT method 
%%% proposed in 
%of .

\subsection{Use case scenario: SPA-based MTT to Fuse AIS Data and SAR Measurements}
\label{subsec:use_case_scenario}
\begin{figure}[!t]
\centering
\includegraphics[width=.5\columnwidth]{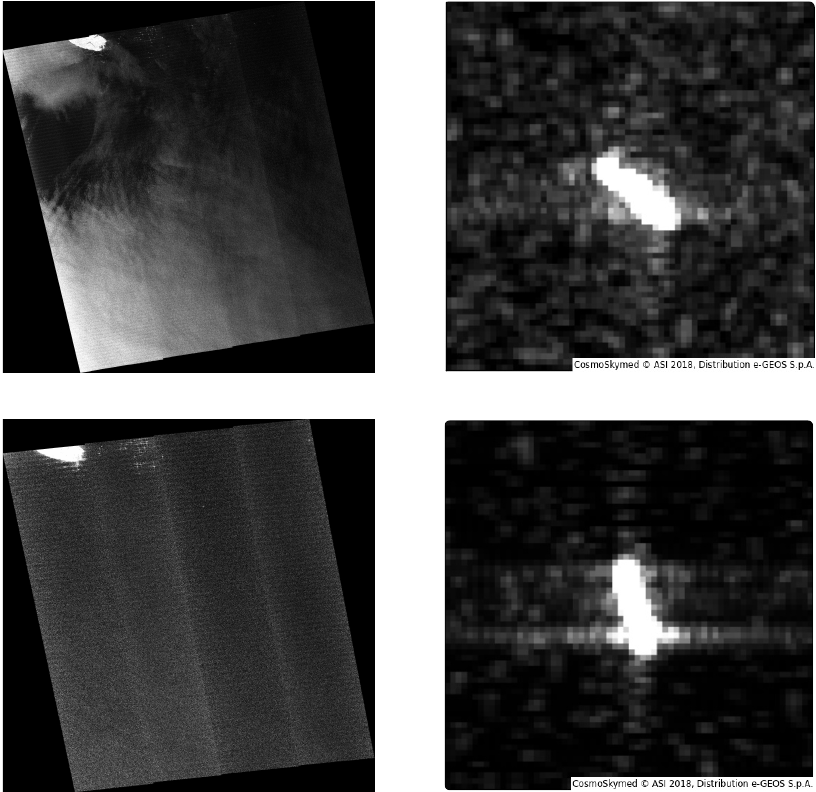}
\caption{(Left images) Subsampled (5x times) quicklook SAR images acquired on 9 September 2018 at 04:37 (Left-Top) and on 10 September 2018 at 04:55 (Left-Bottom) by the COSMO-SkyMed constellation. (Right Images) Full resolution SAR images showing detected ships on 9 September 2018 at 04:37 (Top-Right) and on 10 September 2018 at 04:55 (Bottom-Right). (All the images are of ASI property and processed by E-GEOS.)}
\label{fig:SAR_detections_GAIA_LEADER}
\end{figure}

The use-case scenario described in the current section aims at demonstrating how SPA-based MTT methods can be efficiently used to fuse measurements (ship detections) extracted from SAR images, which do not provide any information regarding ship identities, and the information contained in the AIS messages, i.e., ship's position and the MMSI identifier~\cite{GaglioneBS:C18,SoldiGMHBFW:C19}. The AIS messages and SAR measurements were collected from an area of the Mediterranean Sea, off the shore of Malta, in the period from 8 September 2018 at 16:30 till 12 September 2018 at 16:47. The SAR measurements are extracted from a sequence of SAR images acquired by the COSMO-SkyMed constellation, by means of AI segmentation techniques as described in Section~\ref{subsec:main_architectures_obj_class}, and consist of estimated GPS coordinates of the detected targets, i.e., latitude and longitude. 
The time interval between two consecutive AIS messages is at least of 5 hours. 
Fig.~\ref{fig:SAR_detections_GAIA_LEADER} shows two examples of subsampled quicklook SAR images (left column) acquired by the 
COSMO-SkyMed constellation and two examples of full resolution SAR images (right column) showing detected 
ships. 
Due to the MOU problem, it is not known in advance whether a SAR measurement is originated from a target or is a false alarm, and, in the case it has been originated by a target, it is not known by which target. 
%The time interval between two consecutive AIS measurements is at least
%of 5 hours; these measurements provide besides the information on the ship's position also the MMSI identifier of the ship sending the AIS information.
\begin{figure}[!t]
\centering
%\begin{postscript}
%\centering
%\psfrag{Malta}[c][c][.85]{\raisebox{0mm}{\hspace{.15mm}Malta}}
%\psfrag{xLabel}[c][c][.65]{\raisebox{-6mm}{\hspace{.15mm}Longitude}}
%\psfrag{yLabel}[c][c][.65]{\raisebox{-2mm}{\hspace{.15mm}Latitude}}
%\psfrag{first}[c][c][.65]{\raisebox{-6mm}{\hspace{.15mm}(a)}}
%\psfrag{second}[c][c][.65]{\raisebox{-6mm}{\hspace{.15mm}(b)}}
%\psfrag{x1}[c][c][.65]{\raisebox{0mm}{\hspace{.15mm}$14.5^\circ$ E}}
%\psfrag{x2}[c][c][.65]{\raisebox{0mm}{\hspace{.15mm}$15.0^\circ$ E}}
%\psfrag{x3}[c][c][.65]{\raisebox{0mm}{\hspace{.15mm}$15.5^\circ$ E}}
%\psfrag{y1}[c][c][.65]{\raisebox{0mm}{\hspace{-7mm}$36.0^\circ$ N}}
%\psfrag{y2}[c][c][.65]{\raisebox{0mm}{\hspace{-7mm}$35.5^\circ$ N}}
%\psfrag{SAR}[l][l][.65]{\raisebox{0mm}{\hspace{0mm}SAR measurements}}
%\psfrag{AIS}[l][l][.65]{\raisebox{0mm}{\hspace{0mm}AIS measurements}}
%\psfrag{estTraj}[l][l][.65]{\raisebox{0mm}{\hspace{0mm}Estimated trajectory (GAIA LEADER)}}
%\psfrag{trueTraj}[l][l][.65]{\raisebox{0mm}{\hspace{0mm}Real trajectory (GAIA LEADER) }}
\includegraphics[width=.55\columnwidth]{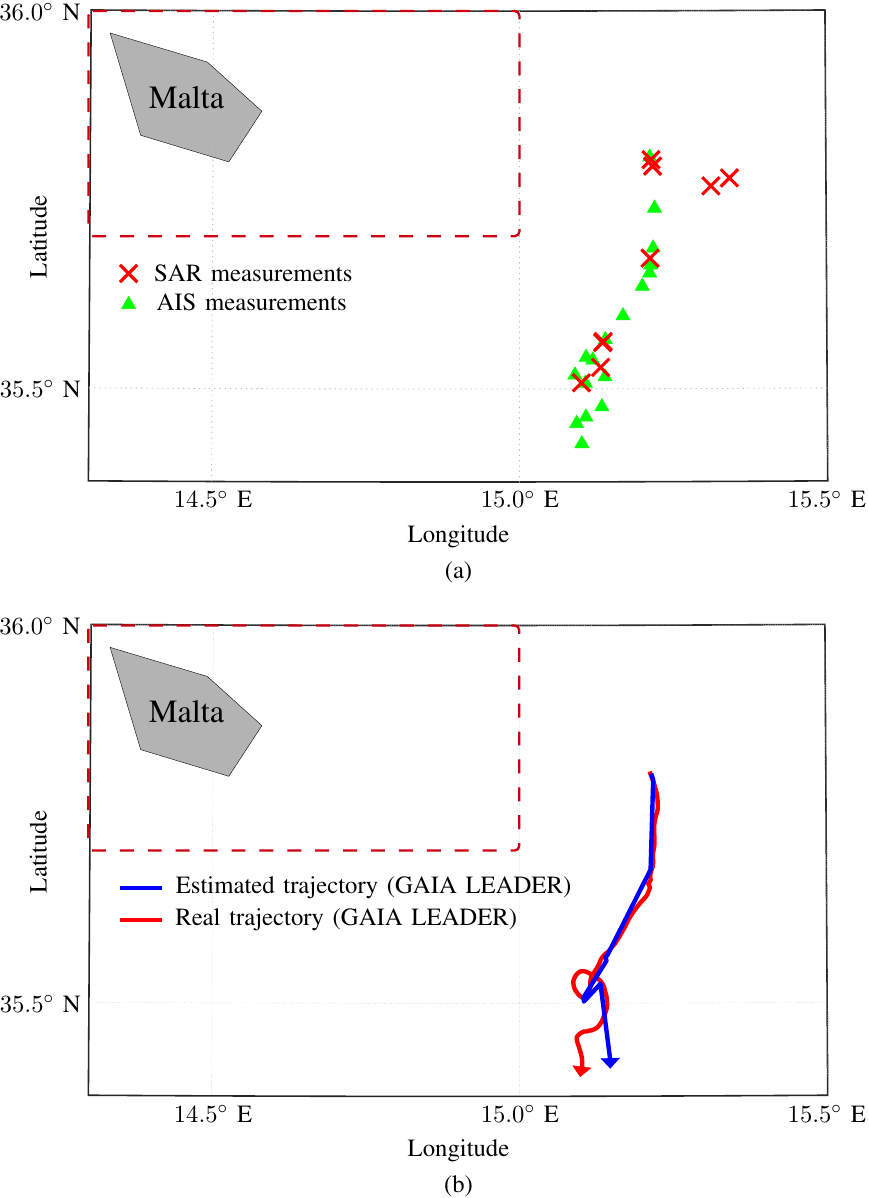}
%\end{postscript}
\caption{(a) SAR (red crosses) and AIS (green triangles) measurements used by the SPA-based fusion algorithm obtained in the period from 8 September 2018 at 16:30 till 12 September 2018 at 16:47. The measurements inside the red dashed rectangle have not  been considered. (b) Estimated and real (derived by AIS messages) trajectories for the vehicles carrier GAIA LEADER obtained using the SAR and AIS measurements.}
\label{fig:useCaseScenarioImage}
\end{figure}
%\begin{figure}[!t]
%\centering
%\includegraphics[width=.5\columnwidth]{./Figures/GAIA_LEADER_2}
%\caption{The vehicles carrier GAIA LEADER. (Image courtesy of MarineTraffic.com)}
%\label{fig:vehicles_cerrier}
%\end{figure}

The main purpose of this use case is therefore to fuse SAR measurements with AIS messages in order to track potential targets and also 
identify them by exploiting the MMSI identifiers. The SAR measurements and AIS messages used by the SPA-based MTT method are represented in Fig.~\ref{fig:useCaseScenarioImage}-(a) by the red crosses and green triangles, respectively. The 
red dashed rectangle delimits a geographical area of little interest from a tracking point of view with a large number of docked ships. Therefore, the SAR measurements and AIS messages from ships within this red dashed rectangle are discarded.
The dynamic model used to describe ship motion in the SPA-based MTT algorithm is the OU model. Estimated positions of detected targets 
are calculated each time a SAR measurement is obtained, using all AIS messages gathered in the time interval between the current 
and the previous SAR measurements. Besides estimating the number of targets and their positions, the algorithm associates, 
when possible, an MMSI identifier to each of them. These MMSI identifiers are selected from a set consisting of all the MMSI 
identifiers so far observed. 
Fig.~\ref{fig:useCaseScenarioImage}-(b) shows the output of the SPA-based MTT algorithm. One can observe that in the considered time 
period and area only a single target, identified by the algorithm as the vehicles carrier GAIA LEADER, is present. In particular,  the red and blue lines show the real (derived by AIS messages) and estimated trajectories of the vehicles carrier GAIA LEADER. %, shown in Fig.~\ref{fig:vehicles_cerrier}. %built in 2011 and sailing under the flag of Panama. 
%Its length overall (LOA) is 199.94 meters and its width is 32.26 meters. 
The SPA-based MTT algorithm is therefore able to provide fused trajectories of the targets using both SAR measurements and AIS messages and allows the identification of each estimated trajectory by associating it to the most likely MMSI identifier.

\section{Conclusions}
\label{sec:conclusion_fut_work}
 
%Maritime situational awareness (MSA) is crucial for search and rescue operations, fishery monitoring, pollution control, law enforcement, and national security policies. The goal of MSA 
%aims to provide 
%%% seamless 
%wide-area operational pictures of ship traffic
%in real time. Most of the terrestrial sensors, such as terrestrial radars or the ground-based Automatic Identification System (AIS), are not able to guarantee a seamless coverage in remote ares of the oceans. However, the increasing spread of satellite constellations orbiting around the Earth and providing continuous Earth observation can be an advantage to enhance MSA even in Earth's areas that would not be otherwise covered by conventional sensors. 
%%In particular, this paper has provided an overview of the main space-based sensor technologies, i.e., satellite AIS (Sat-AIS), Synthetic Aperture Radars (SAR), multi-spectral and hyper-spectral optical sensors and global navigation satellite systems (GNSS), and has presented the advantages and limitations of each technology in the scope of MSA.
The increasing availability of multiple space-based sensors providing detailed images of the ships at sea calls for the development of advanced image analysis techniques for target detection, segmentation and classification. In a companion paper~\cite{SpaceAESM_1:J20}, we introduced and discussed several satellite technologies for Maritime Surveillance (MS). In this paper, we reported the most recent deep-learning techniques and described their effectiveness in several use cases exploiting SAR, high-resolution (HR) and very high-resolution (VHR) images. 
We have presented Bayesian and allied statistical techniques to extract information (for example, common sea-lanes) from repositories of historical Sat-AIS data, and have shown how to track multiple targets using heterogeneous space-based and terrestrial sensors. Multitarget tracking (MTT) approaches based on the sum-product algorithm (SPA) have gained strong popularity thanks to their flexibility and scalability.
A use case scenario that demonstrates the fusion of measurements extracted from SAR images
acquired near the shores of Malta island with AIS messages highlights the effectiveness of the SPA-based MTT methods in the context of MS using space-based sensors. 

\bibliographystyle{IEEEtran}
%\bibliography{Bibliography/IEEEabrv,Bibliography/WGroup,Bibliography/BiblioCV,Bibliography/Temp,Bibliography/Temp_AS,Bibliography/Temp_NF,Bibliography/myBib}
\bibliography{Space-Based-MSA-AESM-V4_Part_2.bbl}

\begin{thebibliography}{10}
\providecommand{\url}[1]{#1}
\csname url@samestyle\endcsname
\providecommand{\newblock}{\relax}
\providecommand{\bibinfo}[2]{#2}
\providecommand{\BIBentrySTDinterwordspacing}{\spaceskip=0pt\relax}
\providecommand{\BIBentryALTinterwordstretchfactor}{4}
\providecommand{\BIBentryALTinterwordspacing}{\spaceskip=\fontdimen2\font plus
\BIBentryALTinterwordstretchfactor\fontdimen3\font minus
  \fontdimen4\font\relax}
\providecommand{\BIBforeignlanguage}[2]{{%
\expandafter\ifx\csname l@#1\endcsname\relax
\typeout{** WARNING: IEEEtran.bst: No hyphenation pattern has been}%
\typeout{** loaded for the language `#1'. Using the pattern for}%
\typeout{** the default language instead.}%
\else
\language=\csname l@#1\endcsname
\fi
#2}}
\providecommand{\BIBdecl}{\relax}
\BIBdecl

\bibitem{SpaceAESM_1:J20}
G.~Soldi, D.~Gaglione, N.~Forti, A.~D. Simone, F.~C. Daffin\`{a}, G.~Bottini,
  D.~Quattrociocchi, L.~M. Millefiori, P.~Braca, S.~Carniel, P.~Willett,
  A.~Iodice, D.~Riccio, and A.~Farina, ``Space-based global maritime
  surveillance. {P}art {I}: {S}atellite technologies,'' \emph{{IEEE} Aerosp.
  Electron. Syst. Mag. {(Submitted)}}, 2020.

\bibitem{GooBenCou:B16}
\BIBentryALTinterwordspacing
I.~Goodfellow, Y.~Bengio, and A.~Courville, ``Deep learning,'' 2016, book in
  preparation for MIT Press. [Online]. Available:
  \url{http://www.deeplearningbook.org}
\BIBentrySTDinterwordspacing

\bibitem{Kendall:J15}
A.~Kendall, V.~Badrinarayanan, , and R.~Cipolla, ``Bayesian segnet: Model
  uncertainty in deep convolutional encoder-decoder architectures for scene
  understanding,'' \emph{arXiv preprint arXiv:1511.02680}, 2015.

\bibitem{Ronnebergem:J15}
O.~Ronneberger, P.~Fischer, and T.~Brox, ``{U-Net}: Convolutional networks for
  biomedical image segmentation,'' in \emph{Medical Image Computing and
  Computer-Assisted Intervention (MICCAI)}, ser. LNCS, vol. 9351.\hskip 1em
  plus 0.5em minus 0.4em\relax Springer, 2015, pp. 234--241.

\bibitem{CapobiancoSM18}
S.~{Capobianco}, L.~{Scommegna}, and S.~{Marinai}, ``Historical handwritten
  document segmentation by using a weighted loss,'' in \emph{Proc. {ANNPR}},
  2018, pp. 395--406.

\bibitem{GargiuloDIRR19}
M.~{Gargiulo}, D.~A.~G. {Dell’Aglio}, A.~{Iodice}, D.~{Riccio}, and
  G.~{Ruello}, ``Semantic segmentation using deep learning: {A} case of study
  in {A}lbufera {P}ark, {V}alencia,'' in \emph{Proc. {IEEE} {MetroAgriFor}},
  2019, pp. 134--138.

\bibitem{HeGkDoGi:C17}
K.~He, G.~Gkioxari, P.~Dollár, and R.~Girshick, ``{Mask R-CNN},'' in
  \emph{2017 IEEE International Conference on Computer Vision (ICCV)}, 2017,
  pp. 2980--2988.

\bibitem{adelson1984}
E.~H. {Adelson}, C.~H. {Anderson}, J.~R. {Bergen}, P.~J. {Burt}, and J.~M.
  {Ogden}, ``{Pyramid methods in image processing},'' \emph{{RCA} {E}ngineer},
  vol.~29, no.~6, pp. 33--41, 1984.

\bibitem{HeZhangRenSun:C16}
K.~{He}, X.~{Zhang}, S.~{Ren}, and J.~{Sun}, ``Deep residual learning for image
  recognition,'' in \emph{2016 IEEE Conference on Computer Vision and Pattern
  Recognition (CVPR)}, 2016, pp. 770--778.

\bibitem{survey19}
Z.~{Xiao}, X.~{Fu}, L.~{Zhang}, and R.~S.~M. {Goh}, ``Traffic pattern mining
  and forecasting technologies in maritime traffic service networks: A
  comprehensive survey,'' \emph{{IEEE} Trans. Intell. Transp. Syst.}, pp.
  1--30, 2019.

\bibitem{Zissis20}
D.~{Zissis}, K.~{Chatzikokolakis}, G.~{Spiliopoulos}, and M.~{Vodas}, ``A
  distributed spatial method for modeling maritime routes,'' \emph{IEEE
  Access}, vol.~8, pp. 47\,556--47\,568, 2020.

\bibitem{Zissis2017}
A.~Valsamis, K.~Tserpes, D.~Zissis, D.~Anagnostopoulos, and T.~Varvarigou,
  ``Employing traditional machine learning algorithms for big data streams
  analysis: The case of object trajectory prediction,'' \emph{Journal of
  Systems and Software}, vol. 127, pp. 249--257, 2017.

\bibitem{Nguyen2018}
D.~Nguyen, R.~Vadaine, G.~Hajduch, R.~Garello, and R.~Fablet, ``A multi-task
  deep learning architecture for maritime surveillance using {AIS} data
  streams,'' \emph{IEEE International Conference on Data Science and Advanced
  Analytics}, pp. 331--340, 2018.

\bibitem{icassp20}
N.~Forti, L.~M. Millefiori, P.~Braca, and P.~Willett, ``Prediction of vessel
  trajectories from {AIS} data via sequence-to-sequence recurrent neural
  networks,'' in \emph{IEEE International Conference on Acoustics, Speech and
  Signal Processing}, 2020.

\bibitem{Cazza16}
L.~{Cazzanti}, A.~{Davoli}, and L.~M. {Millefiori}, ``Automated port traffic
  statistics: From raw data to visualisation,'' in \emph{IEEE International
  Conference on Big Data}, 2016, pp. 1569--1573.

\bibitem{Li16}
Y.~Li, R.~W. Liu, J.~Liu, Y.~Huang, B.~Hu, and K.~Wang, ``Trajectory
  compression-guided visualization of spatio-temporal {AIS} vessel density,''
  in \emph{8th International Conference on Wireless Communications {\&} Signal
  Processing}, 2016.

\bibitem{Bomb06}
N.~A. {Bomberger}, B.~J. {Rhodes}, M.~{Seibert}, and A.~M. {Waxman},
  ``Associative learning of vessel motion patterns for maritime situation
  awareness,'' in \emph{9th International Conference on Information Fusion},
  2006.

\bibitem{Ristic08}
B.~{Ristic}, B.~{La Scala}, M.~{Morelande}, and N.~{Gordon}, ``Statistical
  analysis of motion patterns in {AIS} data: Anomaly detection and motion
  prediction,'' in \emph{International Conference on Information Fusion}, 2008.

\bibitem{Xiao17}
Z.~{Xiao}, L.~{Ponnambalam}, X.~{Fu}, and W.~{Zhang}, ``Maritime traffic
  probabilistic forecasting based on vessels’ waterway patterns and motion
  behaviors,'' \emph{{IEEE} Trans. Intell. Transp. Syst.}, vol.~18, no.~11, pp.
  3122--3134, 2017.

\bibitem{Pallo13}
G.~Pallotta, M.~Vespe, and K.~Bryan, ``Vessel pattern knowledge discovery from
  ais data: A framework for anomaly detection and route prediction,''
  \emph{Entropy}, vol.~15, no.~6, pp. 2218--2245, 2013.

\bibitem{Coscia2018}
P.~{Coscia}, P.~{Braca}, L.~M. {Millefiori}, F.~A.~N. {Palmieri}, and
  P.~{Willett}, ``Multiple {O}rnstein-{U}hlenbeck processes for maritime
  traffic graph representation,'' \emph{{IEEE} Trans. Aerosp. Electron. Syst.},
  vol.~54, no.~5, pp. 2158--2170, 2018.

\bibitem{oceans2019}
N.~Forti, L.~Millefiori, and P.~Braca, ``Unsupervised extraction of maritime
  patterns of life from {Automatic Identification System} data,'' in
  \emph{MTS/IEEE OCEANS}, 2019.

\bibitem{Millefiori2016}
L.~M. Millefiori, P.~Braca, K.~Bryan, and P.~Willett, ``Modeling vessel
  kinematics using a stochastic mean-reverting process for long-term
  prediction,'' \emph{{IEEE} Trans. Aerosp. Electron. Syst.}, vol.~52, no.~5,
  pp. 2313--2330, 2016.

\bibitem{Millefiori2016-2}
L.~M. Millefiori, P.~Braca, and P.~Willett, ``Consistent estimation of randomly
  sampled {O}rnstein-{U}hlenbeck process long-run mean for long-term target
  state prediction,'' \emph{{IEEE} Signal Process. Lett.}, vol.~23, no.~11, pp.
  1562--1566, 2016.

\bibitem{Millefiori2015}
L.~M. {Millefiori}, G.~{Pallotta}, P.~{Braca}, S.~{Horn}, and K.~{Bryan},
  ``Validation of the {O}rnstein-{U}hlenbeck route propagation model in the
  {M}editerranean {S}ea,'' in \emph{MTS/IEEE OCEANS}, 2015.

\bibitem{Millefiori2017}
L.~M. Millefiori, P.~Braca, and G.~Arcieri, ``Scalable distributed change
  detection and its application to maritime traffic,'' in \emph{IEEE
  International Conference on Big Data}, 2017, pp. 1650--1657.

\bibitem{dbscan}
M.~Ester, H.-P. Kriegel, J.~Sander, and X.~Xu, ``A density-based algorithm for
  discovering clusters in large spatial databases with noise,'' in
  \emph{Intenational Conference on Data Mining and Knowledge Discovery},
  vol.~4, 1996, pp. 226--231.

\bibitem{Lane10}
R.~O. Lane, D.~A. Nevell, S.~D. Hayward, and T.~W. Beaney, ``Maritime anomaly
  detection and threat assessment,'' in \emph{13th International Conference on
  Information Fusion}, 2010, pp. 1--8.

\bibitem{Kowalska12}
K.~Kowalska and L.~Peel, ``Maritime anomaly detection using gaussian process
  active learning,'' in \emph{15th International Conference on Information
  Fusion}, 2012, pp. 1164--1171.

\bibitem{Vespe12}
M.~Vespe, I.~Visentini, K.~Bryan, and P.~Braca, ``Unsupervised learning of
  maritime traffic patterns for anomaly detection,'' in \emph{9th IET Data
  Fusion Target Tracking Conference}, 2012, pp. 1--5.

\bibitem{Enrica18}
E.~d'Afflisio, P.~Braca, L.~M. Millefiori, and P.~Willett, ``Detecting
  anomalous deviations from standard maritime routes using the
  {O}rnstein-{U}hlenbeck process,'' \emph{{IEEE} Trans. Signal Process.},
  vol.~66, no.~24, pp. 6474--6487, 2018.

\bibitem{Enrica18-B}
E.~{d'Afflisio}, P.~{Braca}, L.~M. {Millefiori}, and P.~{Willett}, ``Maritime
  anomaly detection based on mean-reverting stochastic processes applied to a
  real-world scenario,'' in \emph{21st International Conference on Information
  Fusion}, 2018, pp. 1171--1177.

\bibitem{fusion18}
N.~{Forti}, L.~M. {Millefiori}, and P.~{Braca}, ``Hybrid {Bernoulli} filtering
  for detection and tracking of anomalous path deviations,'' in \emph{21st
  International Conference on Information Fusion}, 2018, pp. 1178--1184.

\bibitem{icassp19}
N.~{Forti}, L.~M. {Millefiori}, P.~{Braca}, and P.~{Willett}, ``Anomaly
  detection and tracking based on mean--reverting processes with unknown
  parameters,'' in \emph{IEEE International Conference on Acoustics, Speech and
  Signal Processing}, 2019, pp. 8449--8453.

\bibitem{rnc2019}
N.~Forti, G.~Battistelli, L.~Chisci, and B.~Sinopoli, ``Joint attack detection
  and secure state estimation of cyber-physical systems,'' \emph{International
  Journal of Robust and Nonlinear Control}, 2019.

\bibitem{BarWilTia:B11}
Y.~Bar-Shalom, P.~K. Willett, and X.~Tian, \emph{{Tracking and Data Fusion: A
  Handbook of Algorithms}}.\hskip 1em plus 0.5em minus 0.4em\relax Storrs, CT,:
  Yaakov Bar-Shalom, 2011.

\bibitem{VivoneBH15_53}
G.~Vivone, P.~Braca, and J.~Horstmann, ``Knowledge-based multitarget ship
  tracking for {HF} surface wave radar systems,'' \emph{{IEEE} Trans. Geosci.
  Remote Sens.}, vol.~53, no.~7, pp. 3931--3949, Jul. 2015.

\bibitem{MeyBraWilHla:J17}
F.~Meyer, P.~Braca, P.~Willett, and F.~Hlawatsch, ``{A scalable algorithm for
  tracking an unknown number of targets using multiple sensors},'' \emph{{IEEE}
  Trans. Signal Process.}, vol.~65, no.~13, pp. 3478--3493, Jul. 2017.

\bibitem{MeyKroWilLauHlaBraWin:J18}
F.~Meyer, T.~Kropfreiter, J.~L. Williams, R.~A. Lau, F.~Hlawatsch, P.~Braca,
  and M.~Z. Win, ``Message passing algorithms for scalable multitarget
  tracking,'' \emph{Proc. {IEEE}}, vol. 106, no.~2, pp. 221--259, Feb. 2018.

\bibitem{Rei:J79}
D.~B. Reid, ``An algorithm for tracking multiple targets,'' \emph{{IEEE} Trans.
  Autom. Control}, vol.~24, no.~6, pp. 843--854, Dec. 1979.

\bibitem{ChongMR18}
C.~{Chong}, S.~{Mori}, and D.~B. {Reid}, ``Forty years of multiple hypothesis
  tracking - {A} review of key developments,'' in \emph{Proc. FUSION-18}, 2018,
  pp. 452--459.

\bibitem{Mah:B07}
R.~Mahler, \emph{{Statistical Multisource-Multitarget Information
  Fusion}}.\hskip 1em plus 0.5em minus 0.4em\relax Norwood, MA: Artech House,
  2007.

\bibitem{Mah:J03}
R.~P.~S. {Mahler}, ``Multitarget {Bayes} filtering via first-order multitarget
  moments,'' \emph{{IEEE} Trans. Aerosp. Electron. Syst.}, vol.~39, no.~4, pp.
  1152--1178, Oct. 2003.

\bibitem{Mah:J07}
R.~{Mahler}, ``{PHD} filters of higher order in target number,'' \emph{{IEEE}
  Trans. Aerosp. Electron. Syst.}, vol.~43, no.~4, pp. 1523--1543, Oct. 2007.

\bibitem{VoVoCan:J07}
B.-T. Vo, B.-N. Vo, and A.~Cantoni, ``{Analytic implementations of the
  cardinalized probability hypothesis density filter},'' \emph{{IEEE} Trans.
  Signal Process.}, vol.~55, no.~7, pp. 3553--3567, Jul. 2007.

\bibitem{VoVoCan:J09}
------, ``{The cardinality balanced multi-target multi-Bernoulli filter and its
  implementations},'' \emph{{IEEE} Trans. Signal Process.}, vol.~57, no.~2, pp.
  409--423, Feb. 2009.

\bibitem{PonsforW10_18}
A.~Ponsford and J.~Wang, ``A review of high frequency surface wave radar for
  detection and tracking of ships,'' \emph{Turk. J. Elec. Eng. \& Comp. Sci.},
  vol.~18, pp. 409--428, May 2010.

\bibitem{MarescaBHG14_52}
S.~Maresca, P.~Braca, J.~Horstmann, and R.~Grasso, ``Maritime surveillance
  using multiple high-frequency surface-wave radars,'' \emph{{IEEE} Trans.
  Geosci. Remote Sens.}, vol.~52, no.~8, pp. 5056--5071, Aug. 2014.

\bibitem{GagSolMeyHlaBraFarWin:J20}
D.~Gaglione, G.~Soldi, F.~Meyer, F.~Hlawatsch, P.~Braca, A.~Farina, and M.~Z.
  Win, ``Bayesian information fusion and multitarget tracking for maritime
  situational awareness,'' \emph{IET Radar Sonar Navi. (in press)}, 2020.

\bibitem{BarRonKir:01}
Y.~Bar-Shalom, X.~R. Li, and T.~Kirubarajan, \emph{Estimation with Applications
  to Tracking and Navigation}.\hskip 1em plus 0.5em minus 0.4em\relax New York,
  NY: Wiley, 2001.

\bibitem{SoldiB:C18}
G.~Soldi and P.~Braca, ``Online estimation of unknown parameters in
  multisensor-multitarget tracking: a belief propagation approach,'' in
  \emph{Proc. FUSION-18}, Cambridge, U.K., Jul. 2018, pp. 2151--2157.

\bibitem{SoldiMBH:J19}
G.~Soldi, F.~Meyer, P.~Braca, and F.~Hlawatsch, ``Self-tuning algorithms for
  multisensor-multitarget tracking using belief propagation,'' \emph{{IEEE}
  Trans. Signal Process.}, vol.~67, no.~15, pp. 3922--3937, Aug. 2019.

\bibitem{SoldiGMHBFW:C19}
G.~Soldi, D.~Gaglione, F.~Meyer, F.~Hlawatsch, P.~Braca, A.~Farina, and M.~Z.
  Win, ``Heterogeneous information fusion for multitarget tracking using the
  sum-product algorithm,'' in \emph{Proc. IEEE ICASSP-19}, Brighton, U.K., May
  2019, pp. 5471--5475.

\bibitem{GaglioneBS:C18}
D.~Gaglione, P.~Braca, and G.~Soldi, ``Belief propagation based {AIS}/radar
  data fusion for multi-target tracking,'' in \emph{Proc. FUSION-18},
  Cambridge, U.K., Jul. 2018, pp. 2143--2150.

\bibitem{GagSolBraMagMeyHla:J19}
D.~{Gaglione}, G.~{Soldi}, P.~{Braca}, G.~{De Magistris}, F.~{Meyer}, and
  F.~{Hlawatsch}, ``Classification-aided multitarget tracking using the
  sum-product algorithm,'' \emph{{IEEE} Signal Process. Lett.}, vol.~27, pp.
  1710--1714, 2020.

\bibitem{SoldiGDBSFTP:C20}
G.~{Soldi}, D.~{Gaglione}, G.~{De Magistris}, P.~{Braca}, P.~{Stinco},
  G.~{Ferri}, A.~{Tesei}, and K.~{Le Page}, ``Underwater tracking based on the
  sum-product algorithm enhanced by a neural network detections classifier,''
  in \emph{Proc. IEEE ICASSP-20}, May 2019, pp. 5460--5464.

\end{thebibliography}
%\bibliography{../Bibliography/IEEEabrv,../Bibliography/WGroup,../Bibliography/BiblioCV,../Bibliography/Temp}
%\bibliography{./Bibliography/IEEEabrv,./Bibliography/Wgroup,./Bibliography/BiblioCV,./Bibliography/Temp}
\end{document}